\documentclass[%
 preprint,
superscriptaddress,
 amsmath,amssymb, 
endfloats]{revtex4-2}

\usepackage{times}

\usepackage{color}
\usepackage{graphicx}
\usepackage{dcolumn}
\usepackage{bm}
\usepackage{amsmath}
\usepackage{amssymb}
\usepackage{multirow}
\usepackage{tabularx}
\usepackage{ulem}
\usepackage{hyperref}

\newcounter{lastnote}

\begin{document} 

\title{Evidence of Pseudogravitational Distortions of the Fermi Surface Geometry in the Antiferromagnetic Metal FeRh}


\author{Joseph Sklenar}
\affiliation{Department of Physics and Astronomy, Wayne State University, Detroit, MI 48201, USA}
\affiliation{Department of Physics, University of Illinois at Urbana-Champaign, Urbana, IL 61801, USA}
\affiliation{Materials Research Laboratory, University of Illinois at Urbana-Champaign, Urbana, IL 61801, USA}
\author{Soho Shim}
\affiliation{Department of Physics, University of Illinois at Urbana-Champaign, Urbana, IL 61801, USA}
\affiliation{Materials Research Laboratory, University of Illinois at Urbana-Champaign, Urbana, IL 61801, USA}
\author{Hilal Saglam}
\affiliation{Materials Science Division, Argonne National Laboratory, Lemont IL 60439, USA}
\affiliation{Department of Physics, Yale University, New Haven, Connecticut 06520, USA}
\author{Junseok Oh}
\affiliation{Department of Physics, University of Illinois at Urbana-Champaign, Urbana, IL 61801, USA}
\affiliation{Materials Research Laboratory, University of Illinois at Urbana-Champaign, Urbana, IL 61801, USA}
\author{M. 
G. 
Vergniory}
\affiliation{Donostia International Physics Center, P. 
Manuel de Lardizabal 4, 20018 Donostia–San Sebastian, Spain}
\affiliation{Max Planck Institute for Chemical Physics of Solids, Dresden, D-01187, Germany.}
\author{Axel Hoffmann}%
\affiliation{Materials Science Division, Argonne National Laboratory, Lemont IL 60439, USA}
\affiliation{ Department of Materials Science and Engineering, University of Illinois at Urbana-Champaign, Urbana, IL 61801, USA
}%
\affiliation{ Department of Electrical and Computer Engineering, University of Illinois at Urbana-Champaign, Urbana, IL 61801, USA
}%
\affiliation{Materials Research Laboratory, University of Illinois at Urbana-Champaign, Urbana, IL 61801, USA}
\affiliation{Department of Physics, University of Illinois at Urbana-Champaign, Urbana, IL 61801, USA}
\author{Barry Bradlyn}
\affiliation{Department of Physics, University of Illinois at Urbana-Champaign, Urbana, IL 61801, USA}
\affiliation{Institute for Condensed Matter Theory, University of Illinois at Urbana-Champaign, Urbana, IL 61801, USA}
\author{Nadya Mason}
\affiliation{Department of Physics, University of Illinois at Urbana-Champaign, Urbana, IL 61801, USA}
\affiliation{Materials Research Laboratory, University of Illinois at Urbana-Champaign, Urbana, IL 61801, USA}
\author{Matthew J. 
Gilbert}
\affiliation{ 
Department of Electrical and Computer Engineering, University of Illinois at Urbana-Champaign
}%


\date{\today}



\baselineskip=24pt


\maketitle 

\section{Abstract}
\label{sec:abstract}
\textbf{The confluence between high-energy physics and condensed matter has produced groundbreaking results via unexpected connections between the two traditionally disparate areas. In this work, we elucidate additional connectivity between high-energy and condensed matter physics by examining the interplay between spin-orbit interactions and local symmetry-breaking magnetic order in the magnetotransport of thin-film magnetic semimetal FeRh. We show that the change in sign of the normalized longitudinal magnetoresistance observed as a function of increasing in-plane magnetic field results from changes in the Fermi surface morphology. We demonstrate that the geometric distortions in the Fermi surface morphology are more clearly understood via the presence of pseudogravitational fields in the low-energy theory. The pseudogravitational connection provides additional insights into the origins of a ubiquitous phenomenon observed in many common magnetic materials and points to an alternative methodology for understanding phenomena in locally-ordered materials with strong spin-orbit interactions.}       

\section{Introduction}
\label{sec:introduction}

The principles of symmetry and symmetry breaking form one of the cornerstones of our understanding of physics. The advent of topological materials has reinforced this fact, while also highlighting the points where symmetry principles alone cannot provide sufficient understanding of a diverse array of materials such as insulators\cite{hasan2010colloquium, qi2011topological}, superconductors\cite{SatoReview}, semimetals\cite{xu2015discovery, yan2017topological, armitage2018weyl} and magnets\cite{FGBafm2013, vsmejkal2018topological}. Of the multitude of materials that have been shown to harbor topological phases, the interplay between band topology and local order in topological superconductors and topological magnets are of particular interest as it is unclear as to the manner in which symmetry and topology compete to control the observable properties of the underlying material. Magnetic metals possess many beneficial aspects compared to superconductors, that may lead to device applicability, such as high Curie temperatures. The high-temperature magnetic phase permits the observation of the competition between topology and magnetism to be explored at temperatures in excess of room temperature in topological magnets. Within this context, the interplay between topology and magnetic order\cite{MacDonaldReview,PhysRevLett.124.066401,TopologicalAFMReview2021} allows for deeper insights into hitherto under-explored experimental signatures observed in both emergent and well-known materials.

In this work, we provide evidence that the interaction between the local magnetic order and spin-orbit coupling manifest in field-tunable distortions of the Fermi surface geometry in FeRh thin films. We use magnetotransport measurements in conjunction with detailed ab-initio and theoretical analysis to show that the thin-film form of antiferromagnetically-ordered (AFM) metal FeRh is a Weyl metal. Furthermore, we demonstrate that the anisotropic magnetoresistance (AMR), measured as a function of the field angle and current direction, may be understood in terms of changes in the Fermi surface topology brought about by geometric distortions that take the form of a coupling of electrons at the Fermi surface to pseudogravitational fields. In this work, we define "pseudogravitational" to refer to geometric distortions of the Fermi surface brought about by the interplay between the electronic bands and the applied electromagnetic fields that mimic the effects of gravity without possessing the properties inherent in true gravitational fields. We demonstrate through simple models that the theoretical mapping of the distorted Fermi surface to a pseudogravitational metric  allows for a coherent and consistent understanding of magnetotransport in clean spin-orbit coupled metals, both with and without nontrivial topology. One of the crucial limitations in probing topological magnetic metals  is their lack of tunability: typical knobs such as strain, external magnetic field, and chemical substitution are limited in their ability to change topological properties. We argue that one potential remedy to this problem is to exploit the interplay between spin-orbit coupling and local symmetry breaking order. The onset of magnetic order, for instance, is known to drastically alter band structure, allowing for the realization of additional symmetries that have resulted in the observation of topological phases harbored in previously unexplored metals and insulators\cite{MacDonaldReview,PhysRevLett.124.066401,TopologicalAFMReview2021,FGBafm2013, vsmejkal2018topological}. 

\section{Results and Discussion}

\subsection{Magnetic and Electronic structure of Antiferromagnetic $FeRh$}
\label{sec:tchar}

\begin{figure*}
\includegraphics[width=\textwidth]{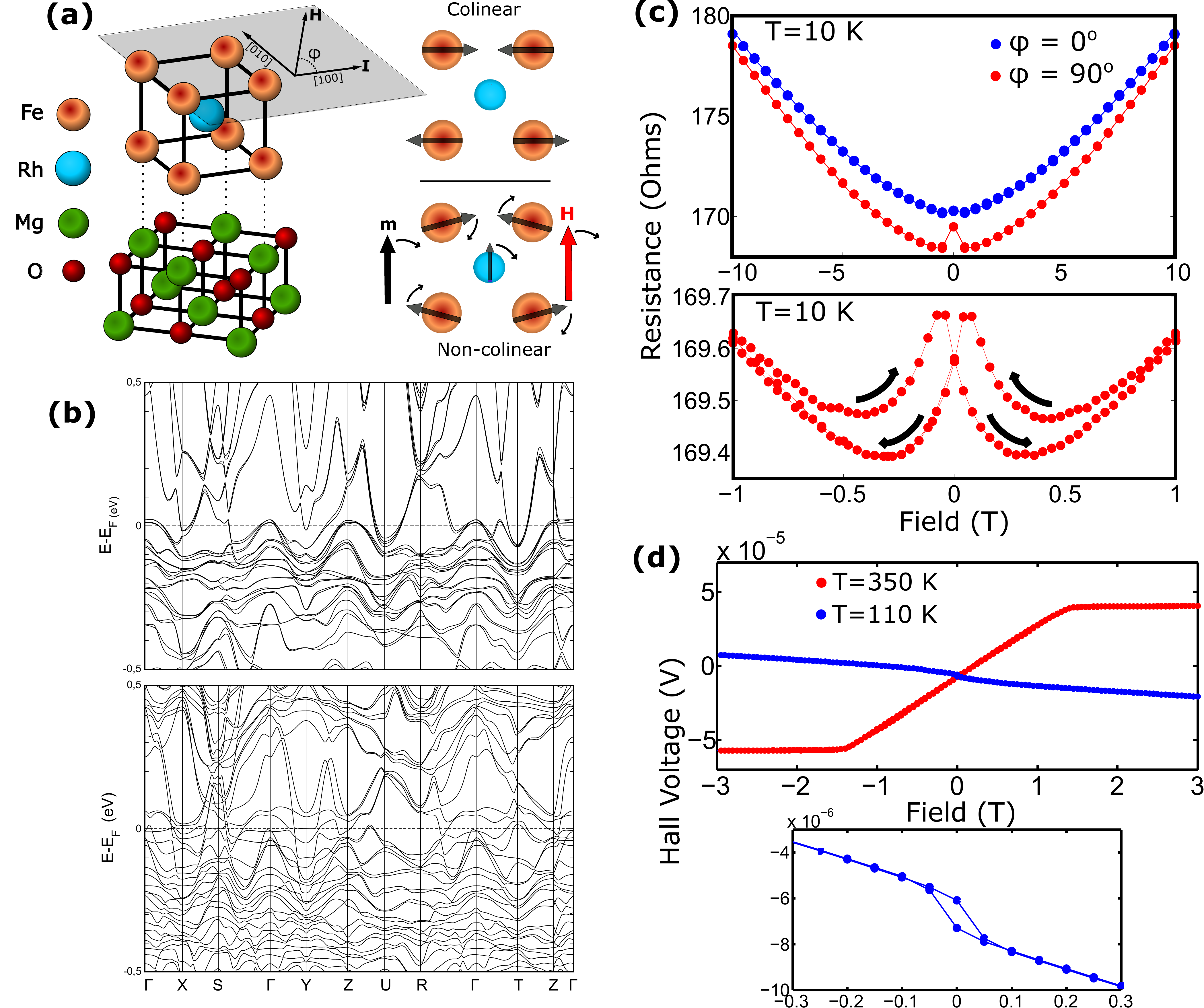}
\caption{\textbf{Characterization and Magnetotransport in Antiferromagnetic Thin-Film FeRh:} (a) Schematic of the FeRh lattice with epitaxial matching to an MgO substrate. $\phi$ is the orientation of the external magnetic field, \textbf{H}, relative to the current along a well-defined crystallographic direction. The behavior of the antiferromagnetic (AFM) order is displayed in the limits of both low and high magnetic fields. An increasing field cants the AFM moments into a non-collinear configuration, and a ferromagnetic moment is generated on the Rh sublattice. The rotation of all magnetization orientations is illustrated as \textbf{H} rotates. Note that the theoretically expected orthorhombic distortion of the FeRh is not indicated in the schematic. (b) Electronic band structure calculations from density functional theory are shown for the orthorhombic collinear AFM  structure (top panel) as well as the non-collinear AFM structure with Rh moment oriented along the $[100]$ direction (bottom panel). (c) The field-dependent magnetoresistance is shown when the field is swept at $\phi = 0^\circ$ and $\phi = 90^\circ$ at T = $10$ K. The magnetoresistance does not saturate in high-magnetic fields, and the lower plot demonstrates a hysteretic anisotropic magnetoresistance peak at low fields consistent with the presence of residual ferromagnetism. The black arrows denote the direction of the external field sweep. (d) The anomalous Hall effect in both the ferro- and antiferromagnetic phase of FeRh are shown at T = $350$ K and $110$ K respectively. The zero-field anomalous voltage, better seen in the lower plot, indicates the presence of a Berry phase induced by strong spin-orbit coupling the material and is concomitant with the presence of a topological response\cite{huang2015observation, nayak2016large,liu2018giant,nakatsuji2015large}.}
\label{fig:100_Data}
\end{figure*}

The FeRh films we examine are sputter-deposited onto a $[001]$-oriented MgO substrate\cite{saglam2019spin}. Epitaxial growth occurs such that the $[100]$-direction of the FeRh grows along the $[110]$-direction of the MgO, as illustrated in Fig.~\ref{fig:100_Data}(a). Although bulk FeRh is cubic, biaxial strain from the lattice mismatch with the substrate causes a small lattice distortion, such that the lattice constant of FeRh along the $[001]$-direction is approximately $2\%$ larger than the in-plane lattice constant\cite{fan2010ferromagnetism}. Furthermore, first-principles calculations suggest that in the antiferromagnetic phase the crystal is additionally orthorhombically distorted in the plane of the MgO\cite{zarkevich2018ferh}. Complete details on the film growth and structural characterization are found in the Methods, and Supplementary Note 1 of the Supplemental Materials (SM) respectively. 

In bulk FeRh, a metamagnetic phase transition between a ferromagnetic phase at high-temperatures and an antiferromagnetic phase at low-temperatures exists near room temperature\cite{lewis2016coupled}. Thinner films tend to have a depressed transition temperature\cite{han2013suppression}. At low magnetic fields, or fields much less than the exchange energy, our $20$~nm thin films have a metamagnetic transition temperature of approximately $290$~K. Experimental verification of the transition may be found in Supplementary Note 1. To understand the nature of the thin-film FeRh metamagnetic transition, we examine the band structure under the application of an in-plane magnetic field using density functional theory (DFT). At zero field in the AFM phase, the magnetic moments on the Fe atoms are taken to be collinear, and there is no magnetic moment on the Rh atoms. By applying an external field, a canted non-collinear AFM spin structure is induced and the Rh atoms develop a ferromagnetic moment in the direction of the applied field, as illustrated Fig.~\ref{fig:100_Data}(a). In Fig.~\ref{fig:100_Data}(b), we show the resultant DFT calculations of the electronic band structure of thin-film FeRh corresponding to the collinear (top) and non-collinear (bottom) spin structures. The collinear magnetic structure has the symmetries of magnetic space group $P_b2/m$, where we have accounted for the orthorhombic distortion of the thin film form of FeRh\cite{zarkevich2018ferh}. Examining the DFT results, we observe several large electron pockets at the Fermi level throughout the Brillouin zone. The non-collinear magnetic structure shown in Fig.~\ref{fig:100_Data}(b) corresponds to a magnetic field that is oriented along the $[100]$-direction, as indicated in Fig.~\ref{fig:100_Data}(a). Within the DFT calculations, the applied in-plane magnetic field is modeled through a non-zero ferromagnetic Rh moment that points in the same direction as the applied magnetic field. In the presence of the non-zero Rh moment, we observe significant Fermi surface reconstruction, leading to the emergence of topological nodes close to the Fermi level. The non-collinear structure has no orientation-reversing symmetries, and so allows for topologically charged Weyl fermions at generic points in the Brillouin zone; the Weyl fermions seen in Fig.~\ref{fig:100_Data}(b) are pinned to high-symmetry lines in the Brillouin zone by twofold rotational symmetry. We utilize the observations obtained via the DFT calculations of thin-film FeRh to precondition the $8$-band tight-binding model parameters that are  discussed in Sec.~\ref{sec:AMRmain} and is explored in more detail in Supplementary Note 3 of the SM.

For magnetotransport measurements, the films are patterned into Hall bars using a photolithographic and ion-milling process. In all transport measurements, the relative angle between the in-plane magnetic field and the applied current is denoted by $\phi$. Representative field-dependent magnetoresistance effects in the AFM phase of FeRh are shown for $\phi = 0^\circ$ and $90^\circ$ in Fig.~\ref{fig:100_Data}(c). A ferromagnetic contribution to the magnetoresistance in Fig.~\ref{fig:100_Data}(c) is evident as a small peak when the field is swept at $\phi = 90^\circ$. The peak arises from ferromagnetic AMR of the FeRh near the MgO interface as the magnetization rotates from being parallel to perpendicular to the current\cite{jaccard2000uniform}. Hysteresis in the magnetoresistance disappears near $1$~T, and is consistent with magnetometry data that shows the residual ferromagnetic moment saturating at a field magnitude below $1$ T. We observe no evidence of Landau levels or Shubnikov–de Haas oscillations in the non-saturating magnetoresistance, signifying a quenched orbital angular momentum in FeRh. 

\subsubsection{Tight-Binding Model}
\label{sec:AMRmain}
To understand the magnetotransport measurements, we construct a minimal tight-binding model that captures the essential features of the strained thin-film FeRh lattice. We start with primitive lattice vectors are $a_{1}=(100)$, $a_{2}=(010)$, and $a_{3}=(001)$. Although these lattice vectors are cubic, we allow for terms in the tight-binding model that break the cubic symmetry. To completely capture the orthorhombic structure of the FeRh lattice, the unit cell consists of $8$ Fe and $8$ Rh atoms. However, to arrive at a qualitatively accurate model of the bands near the Fermi level, we retain only $2$ Fe and $2$ Rh atoms per unit cell. 

The general form of the Hamiltonian is 
\begin{equation}
\begin{split}
 \mathrm{H}  = \sum\limits_{\langle \langle i,j \rangle \rangle} t_{ij} c^{\dagger}_{i} c_{j} + \sum \limits_{\langle \langle i,j \rangle \rangle, {\langle l \rangle}} c^{\dagger}_{i} i\vec{\lambda_{ij}} \cdot \Vec{\bf{\sigma}} c_{j} \\
 + \Delta \sum\limits_{i} \xi c^{\dagger}_{i} ( \Vec{{\bf{m}}} \cdot \Vec{\bf{\sigma}}) c_i,
 \end{split}
 \label{eq:hamtb}
\end{equation}
where, $c_{i} = (c_{i\uparrow}, c_{i\downarrow})^{T}$ are the electron annihilation operator at site $i$ located at real space point $\mathbf{r}_{i}$ and $\mathbf{\sigma} =  (\sigma_{x}, \sigma_{y}, \sigma_{z})$ represent the Pauli matrices acting on spin, with $\sigma_{0}=\mathbb{I}_{2\times2}$. The Hamiltonian contains three distinct types of terms: the first sum in Eq.~\eqref{eq:hamtb} are the spin-independent hopping terms, where ${\langle ij \rangle}$ indicates the hopping range under consideration and $t_{ij}$ is the corresponding hopping amplitude. The second sum gives the  momentum dependent spin-orbit coupling terms with magnitudes $\vec{\lambda_{ij}}$. In Eq.~\eqref{eq:hamtb}, we have used one $s$-orbital per atomic site; we justify the approximation as: (1) the Rh atoms are not guaranteed to sit perfectly centered within the Fe atoms in strained FeRh and (2) the orbital structure we utilize provides a sufficient approximation to the projected $d$-orbital structure in FeRh at half-filling obtained by integrating out the localized orbital moments\cite{khan1981origin,Hasegawa1987}.

The final term in the Hamiltonian is the AFM exchange term\cite{TangAFM2016} that acts on-site and directly competes with the spin-orbit interactions. The AFM interaction is characterized by the magnitude of the exchange, $\Delta$, and the net magnetization orientation, $\Vec{\bf{m}}$, as illustrated in Fig.~\ref{fig:100_Data}(a). The form of the AFM exchange term is general in nature and is enforced by the presence of the matrix, $\xi$, that ensures that the sign of the exchange is opposite between successive Fe atoms and zero on the Rh atoms (see Supplementary Note 3, Sec. E3). The AFM exchange term is applicable in the presence of an inhomogeneous interaction, such as the Dzyaloshinskii-Moriya interaction, that may result from the interplay between the spin-orbit interaction and the localized, on-site exchange interactions in FeRh\cite{Bogdanov2002}. 

To ensure that our reduced FeRh unit cell adequately captures the band crossings at the Fermi level, we constructed an analogous tight-binding model using the full $16$ atom unit cell and inspected the band structure as the intracell hopping terms are tuned to arrive at four decoupled $4$ atom unit cells. We find that the reduced symmetry of the $4$ atom unit cell only affects the energy bands at higher energies than those considered in this work (See Supplementary Note 3 for more details). 

We experimentally justify the form of the spin-orbit coupling in Eq.~\ref{eq:hamtb} using Fig.~\ref{fig:100_Data}(d), which shows the measured anomalous Hall conductivity (AHC) of the thin-film FeRh devices. The AHC is non-vanishing for both the ferro- and antiferromagnetic phases, demonstrating the presence of strong spin-orbit coupling in FeRh. The anomalous Hall conductivity changes sign in the AFM phase relative to the ferromagnetic phase and, taking into account the residual ferromagnetism at the MgO interface, represents a lower estimate of the anomalous Hall conductivity that is found to be $\sigma_{xy} \approx 10^4$  $\Omega^{-1}$m$^{-1}$. For these measurements the applied magnetic field is out-of-plane in the $[001]$-direction and the observation of a non-zero AHC in the transport characteristics indicates the presence of Berry curvature in the band structure of FeRh. In other studies of similar non-collinear antiferromagnets, the existence of Berry curvature has been predicted\cite{chen2014anomalous} and experimentally confirmed\cite{nakatsuji2015large,nayak2016large} via the generation of non-vanishing anomalous Hall voltages. The non-vanishing anomalous voltage at zero-field is attributed to a ``remnant'' non-collinear configuration that originates from exchange coupling with any remnant residual ferromagnetism on the Rh atoms. Furthermore, recent experiments demonstrate the presence of the inverse spin Hall effect in AFM FeRh, which adds additional support to the presence of strong spin-orbit coupling in this thin-film FeRh\cite{wang2020spin}.

\subsection{$FeRh$ Anisotropic Magnetoresistance: Current in the $[100]$ Crystal Direction }\label{sec:100amr}

\subsubsection{Experimental Results} 

\begin{figure*}
\begin{centering}
\includegraphics[width=\textwidth]{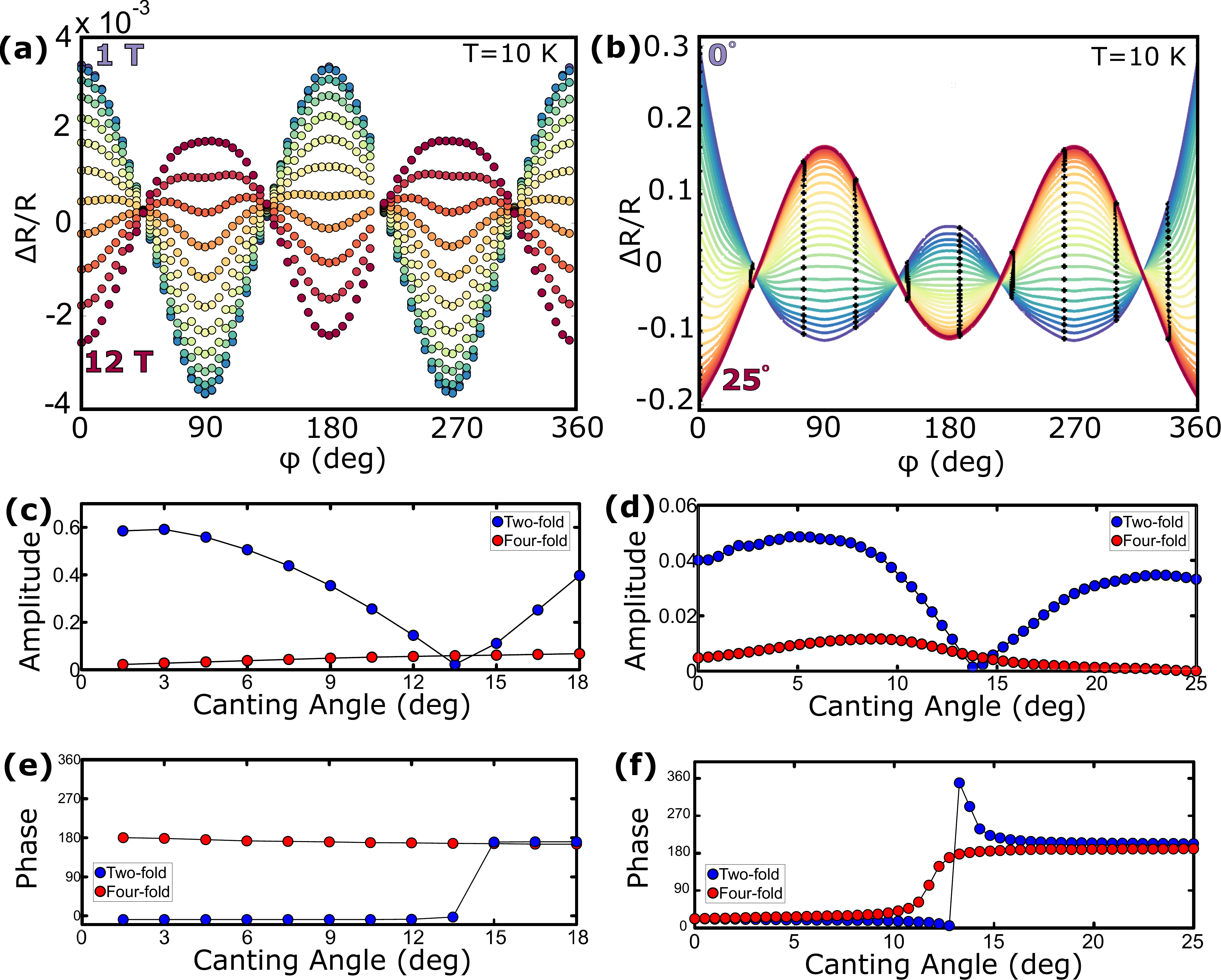}
\caption{\textbf{Quantum Transport in the [100] Crystal Direction:}  (a) Experimentally measured anisotropic magnetoresistance (AMR) for in-plane magnetic fields between $1$-$12$ $T$ We see a clear evolution from two-fold $(C_{2})$ symmetric AMR at low magnetic fields to a four-fold $(C_{4})$, or non-zero coefficient multiplying the $\cos^{4}{\phi}$ term in the AMR spectral decomposition, symmetric AMR and returning to an inverted $C_{2}$ symmetric AMR at high magnetic fields. (b) Theoretically calculated AMR corresponding to canting angles between the Fe moments that lie in the range of $\theta_{cant} = 0^\circ$-$25^\circ$. The spectral decomposition of (c) the experimental and (d) the theoretical AMR amplitudes. In (c), we have converted the experimental field to a canting angle by using an assumed magnetic susceptibility for the thin-film samples that is a factor of three greater than the measured bulk magnetic susceptibility of FeRh. The spectral phase of the harmonics, as a function of the canting angle, for both the experimental and theoretical  AMR measurements are shown in (e) and (f). There is a sharp phase change of $180^{\circ}$ in the experimental $C_{2}$ harmonic that is accurately captured in the theory at the canting angle where the sign change in the AMR is observed but qualitative differences in the $C_{4}$ harmonics exist due to the simplistic nature of the model.}
\label{fig:spectral1}
\end{centering}
\end{figure*}

In Fig.~\ref{fig:spectral1}(a), we show the experimental angular dependence of the AMR for in-plane magnetic fields ranging from $1~T$ to $12~T$, when the current is along the $[100]$ direction of FeRh. We plot the AMR as $\Delta R/R_{avg}$, where $\Delta R = R(\phi) - R_{avg}$. Here, $R(\phi)$ is the resistance when the magnetic field is at an angle $\phi$ relative to the $[100]$ current direction, and $R_{avg}=\frac{1}{2\pi}\int_0^{2\pi}R(\phi)d\phi$ is the average resistance over the entire angular range. As previously shown, smooth AMR signals with a continuous derivative with respect to $\phi$, arise when the magnetic order parameter continuously rotates with the external magnetic field \cite{oh2019angular,baldrati2018full, cheng2019anisotropic, fischer2020large}. Therefore, the continuous AMR we observe in FeRh implies that the ($x-y$) plane is an easy-plane, consistent with previous experimental studies of FeRh\cite{marti2014room}.   

The most striking feature of the AMR is the evolution at fixed $\phi$ as the magnitude of the in-plane external magnetic field increases. Focusing on $\phi=0^{\circ}$ for concreteness, the amplitude of the AMR initially decreases linearly as the magnetic field increases from $1~T$ to $8~T$, where $\Delta R(\phi=0)=0$ at a critical value $B_*\approx 8~T$ of the magnetic field. For fields above $8~T$, the AMR becomes negative and its magnitude increases linearly with increasing magnetic field over the range from $8~T$ to $12~T$. Similar sign changes in the AMR component have been observed in both ferromagnetic Co$_x$Fe$_{1-x}$\cite{PhysRevLett.125.097201}, as well as in antiferromagnetic materials such as EuTiO$_3$\cite{ahadi2019anisotropic}, and Sr$_2$IrO$_4$\cite{wang2014anisotropic}. By estimating the magnetic susceptibility of the thin-film geometry, we may map the critical magnetic field at which the sign of the AMR changes, $B_{c}$,  to a canting angle of $\theta_{cant}\approx13^{\circ}$ for the Fe moments. The canting angle, illustrated in Fig.~\ref{fig:100_Data}(a), is calculated assuming a magnetic susceptibility of the thin-film FeRh that is three times greater than previously measured in bulk FeRh\cite{navarro1999grain}. We note that, due to a strong diamagnetic background signal from the MgO substrate, direct measurement of the magnetic susceptibility is not possible for the thin-film devices used in this work. We expect the presence of an enhanced susceptibility due to the residual ferromagnetism in the FeRh thin-film, near the MgO interface that provides an additional exchange field that aligns with the external magnetic field. 

To quantify the composite amplitudes of each AMR measurement, we perform a spectral analysis of the AMR, writing
\begin{equation}
\Delta R/R_{avg} = \sum_n C_{2n}(B)\cos(2n\phi + \varphi_{2n}).
\end{equation}
The spectral amplitudes $C_{2n}$ of the experimental AMR are shown in Fig.~\ref{fig:spectral1}(c) with corresponding spectral phases $\phi_{2n}$ shown in Fig.~\ref{fig:spectral1}(e). We observe that for fields far from $B_{c}$ the AMR is dominated by the two-fold component $C_{2}$. We also see that $C_{2}(B_{c})=0$, leading to the dominance of the four-fold harmonic $C_{4}$ for fields near $B_{c}$.

\subsubsection{Theoretical Modeling}\label{sec:100model}

To understand the AMR results in thin-film FeRh, we next use our $8$-band tight-binding model to calculate the AMR for current along the $[100]$ crystal direction at T = $10$~K. We utilize the tight-binding Hamiltonian in conjunction with the non-equilibrium Green's function formalism to calculate the observables presented in our work (see Supplementary Note 3 for complete details of the model and Supplementary Table I for the parameter values). As we do not experimentally observe the formation of Landau levels in FeRh, we ignore the orbital effects of the magnetic field. Instead, the external in-plane magnetic field is accounted for by a linear reduction in the antiferromagnetic exchange coupling, $\Delta$ and a corresponding linear change in the cant angle of the local moments from $0^{\circ}-25^{\circ}$\cite{takahashi2002magnetic}. As shown in Fig.~\ref{fig:spectral1}(b), we find that the calculated AMR qualitatively captures all of the essential experimentally observed features, including a sign change in $\Delta R(\phi=0)$, and a transition from two-fold to four-fold to two-fold symmetry as the field passes through a critical value. To facilitate a direct comparison with the experimental data, in Fig.~\ref{fig:spectral1}(d) and (f), we plot the spectral amplitudes $C_{2n}$ and phases $\varphi_{2n}$ of the corresponding theoretical data. By comparing Fig.~\ref{fig:spectral1}(c) and (d), we observe clear qualitative consistencies between theory and experiment. Most notably, we see in the theoretical calculation that the AMR is clearly dominated at all field levels by the twofold symmetric component $C_{2}$ for small cant angles until $\theta_{cant}\approx 13.0^{\circ}$. Furthermore, the sign change of $\Delta R(\phi=0)$ appears in the spectral decomposition as a change in the phase $\phi_{2n}$ by $180^\circ$. We observe this at $\theta^{C_2}_{cant}=14.5^\circ$ in Fig.~\ref{fig:spectral1}(e) for the theoretical model, and at $\theta^{C_2}_{cant}=13^\circ$ in the experimental data in Fig.~\ref{fig:spectral1}(f).

Moreover, the four-fold symmetric signal (determined by $C_{4}$) dominates the observed AMR in FeRh when $C_{2}$ component vanishes in both the theoretical and experimental curves. The residual harmonic content, which is small in magnitude both experimentally and theoretically, is dominated by $C_{4}$. The theoretical trend predicts a sign change that is not observed in the experiments at the magnetic fields considered. We attribute this discrepancy to quasiparticle relaxation effects that were not considered in the model, but that naturally occur in the experimental system. Although a full treatment of a disordered quantum transport calculation is beyond the scope of this work, an analysis of random variations of uncorrelated magnetization domains shows that they do not open gaps in the band structure, and hence result in only small changes to the self-energy. While the presence of uncorrelated magnetic domains is insufficient to open a gap in the bands, the overall resistance of the material will increase; however, we expect this effect to average out in $\Delta R / R_{avg}$. Therefore, the quantum transport calculations of the AMR are insensitive to random magnetic disorder\cite{Kim2018}. 

\subsection{Origin of the AMR: Order Parameter Induced Fermi Surface Deformations}
\label{sec:gravity}

\begin{figure*}
\begin{centering}
\includegraphics[width=\textwidth]{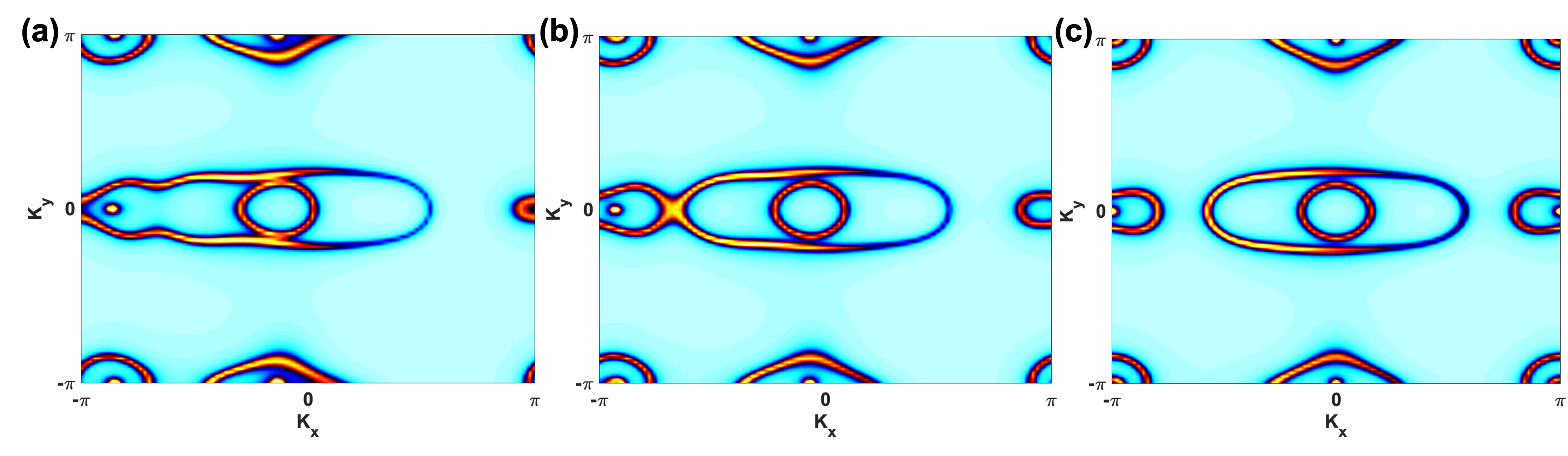}
\caption{\textbf{Spectral Density of States in FeRh along the [100]:} The spectral density of states (SDOS) when the net magnetization orientation, $\hat{m}$ is pointing orthogonal to the different crystal directions in FeRh under three distinct conditions: (i) $\Delta \gg \lambda$ (ii) $\Delta \approx \lambda$ and (iii) $\Delta \ll \lambda$. The rows of plots represent the three conditions in the crystal grouped according to the net magnetization orientation. In the each horizontal row, we plot the SDOS for each of these conditions respectively from the left to the right for the $[100]$ (a)-(c). The SDOS is plotted in the $k_{x}-k_{y}$-plane at $k_{z} = 0$. Each of the three cases illustrates a different point in the Lifshitz transition that occurs in FeRh as the in-plane magnetic field is increased. Furthermore, with each successive increase in magnetic field, the pseudogravitational distortion of the Fermi surface geometry increases.}
\label{fig:LDOS3P}
\end{centering}
\end{figure*}

The $8$-band model of thin-film FeRh allows us to qualitatively reproduce the observed AMR along the $[100]$ direction. 
Going further, our model may be utilized to understand the physical origin of the sign change in the AMR, as well as the relative magnitude of $C_{2}$ and $C_{4}$ as a function of in-plane magnetic field. To do so, we start by examining the local density of states in our tight binding model in three parameter regimes, corresponding to low ($\theta_{cant}<12^{\circ}$, $\Delta \gg \lambda$), intermediate ($12^{\circ}\leq\theta_{cant}\leq 15^{\circ}$, $\Delta \approx \lambda$) and high cant angles ($\theta_{cant}>15^{\circ}$, $\Delta \ll \lambda$). In Fig.~\ref{fig:LDOS3P}, we examine each of these three distinct regimes when the magnetic field is oriented perpendicular to the current direction (i.e.~$\phi=90^\circ$) 
We show the spectral density (SDOS) for the $[100]$ current direction in Fig.~\ref{fig:LDOS3P}(a)  when the ($\theta_{cant}=5^{\circ}$) there exist clear demarcations of the Fermi surfaces that pass through the constant $E_{F}=0~eV$ cut at $k_{z} = 0$ within the $k_{x}-k_{y}$-plane that are connected and reflected about $k_{x}=k_{y}=0$.

Next we increase the cant angle to $\theta_{cant}^{100}=14^{\circ}$, as occurs in the experimental measurements via the application of increasing in-plane magnetic fields. In Fig.~\ref{fig:LDOS3P}(b), we show that this causes the Fermi surfaces to elongate along the $[100]$ direction perpendicular to the $[010]$-directed external field. By continuing to increase the cant angle in Fig.~\ref{fig:LDOS3P}(c)to be $\theta_{cant}^{100}=25^{\circ}$, corresponding to the point when $C_{2}=0$. In each case the formerly concentric Fermi surfaces become disconnected. Crucially, we note that the Fermi contours touch at the same point in parameter space where $C_{2}=0$. The Lifshitz transition in the Fermi surface geometry is thus correlated with the sign change in the AMR. Finally, as the cant angle continues to increase under concomitant increases in the external magnetic field, the Fermi surfaces illustrated in the SDOS of Fig.~\ref{fig:LDOS3P}, become distorted in the direction perpendicular to the external magnetic field, resulting in the observed AMR where $C_{2} < 0$. 

\begin{figure*}
\begin{centering}
\includegraphics[width=\textwidth]{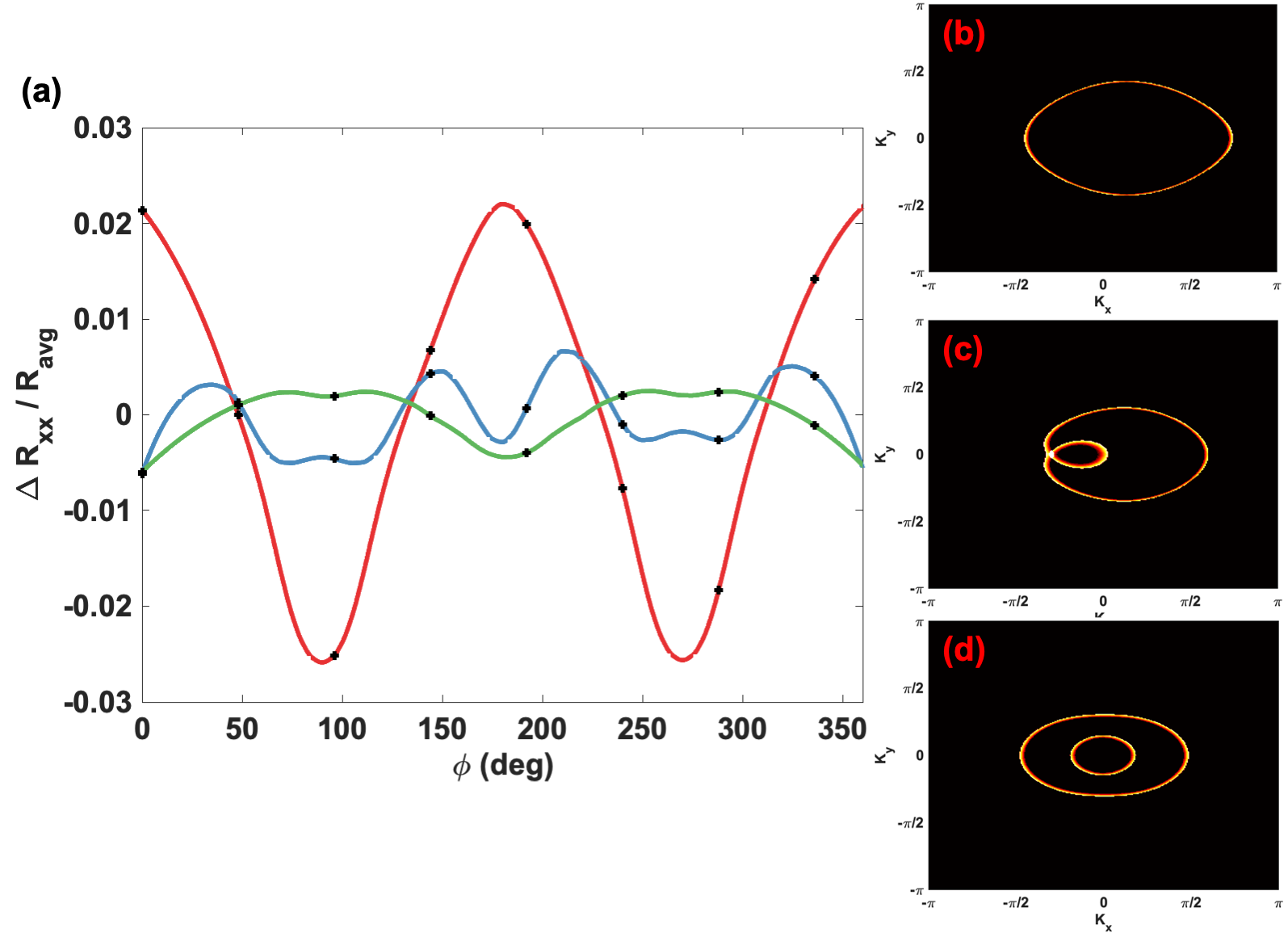}
\caption{\textbf{Quantum Transport in a Rashba Spin-Orbit Coupled Ferromagnet:} (a) Numerically calculated anisotropic magnetoresistance (AMR) corresponding to a spin orbit coupling of $\lambda_{R}=0.5$, a chemical potential of $\mu=0.2$, and $\Delta=0.9$ (red), $\Delta=0.5$ (blue) and $\Delta = 0.01$ (green) where in each case the magnetic order parameter is oriented along the $\hat{y}$-direction in the model. (b) The distorted non-degenerate Fermi surface corresponding to the $C_{2}$ symmetric AMR of $\Delta=0.9$ (c) The Fermi surface when $\Delta=0.5$ shows the appearance of a second concentric Fermi surface at the Fermi level that is indicative of the onset of a Lifshitz transition resulting in the appearance of harmonics beyond that of $C_{2}$. (d) The Fermi surface corresponding to $\Delta=0.01$ where the Fermi level crosses both spin bands resulting in concentric spin-degenerate Fermi surfaces and the recovery of an inverted $C_{2}$-symmetric AMR, as seen in FeRh.}
\label{fig:rashba2b}
\end{centering}
\end{figure*}

Using the $8$-band model for FeRh, we observe that there is a correlation between the change in the sign of $C_{2}$ and a change in geometry and topology of the Fermi surface that occurs as the cant angle is increased. The appearance of a Lifshitz transition as the cant angle, or the ratio of exchange energy to spin-orbit interaction, changes is indicative of a deep connection between the geometry of the Fermi surface and the magnetic order parameter that manifest in the change in sign of the AMR. To clarify the underlying physics, we now construct a minimal model that contains the crucial physical attributes associated with the Lifshitz transition, spin-orbit coupling and the magnetic exchange interaction. To this end, we examine a $2$-band ferromagnetic Rashba spin-orbit coupled metal in two dimensions. The Hamiltonian for the the ferromagnetic metal is
\begin{equation}
H_{FM} = \sum_{\sigma,\langle ij\rangle} -tc^{\dag}_{i\sigma}c_{j\sigma} + i\lambda_Rc^{\dag}_{i\sigma}\hat{z}\times\vec{\sigma}^{\sigma\sigma'}c_{j\sigma'} + \sum_ic^{\dag}_{i\sigma}\vec{\Delta}\cdot\vec{\sigma}^{\sigma\sigma'}c_{j\sigma'}+\sum_{\sigma,i}(4t-\mu)c^{\dag}_{i\sigma} c_{i\sigma},\label{eq:rashbaham}
\end{equation}
where $t$ is the nearest-neighbor hopping amplitude, $\lambda_{R}$ is the Rashba spin-orbit coupling strength, $\vec{\Delta}$ is the ferromagnetic order parameter that may be manipulated in the same manner as Eq.~(\ref{eq:hamtb}), and $\mu$ is the chemical potential. Note, however, that unlike in FeRh, here the value of $|\Delta|$ increases with the external magnetic field due to the ferromagnetic nature of the pairing. In the context of the model, we note that $\sigma$ need not represent the physical spin degree of freedom, instead it may be a spin-orbit coupled degree of freedom that is projected into a set of low-energy bands, as in an antiferromagnet. In Fig.~\ref{fig:rashba2b}(a), we utilize the ferromagnetic Rashba model to calculate the AMR as we rotate the magnetic order parameter. This allows us to emulate the physics of FeRh in the three critical regions surrounding the sign change of $C_{2}$, and thus to codify the interaction between magnetism, Fermi surface geometry, and the spin-orbit interaction. We observe that the three AMR curves faithfully reproduce the intricate physics observed in the more complex tight-binding model for FeRh.

First, we consider the limit $\Delta \gg \lambda_{R}$, where the chemical potential, $\mu$, cuts across only one of the two spin-split bands. This results in a spin non-degenerate Fermi surface, as shown in Fig.~\ref{fig:rashba2b}(b). 
In exploring the Fermi surface, we position the magnetic order parameter along the $\hat{y}$-direction and find the same distortion of the Fermi surface perpendicular to the direction of the order parameter as in our model of FeRh. The result of having a single non-degenerate Fermi surface at the Fermi energy is a two-fold symmetric AMR with a maximum at $\phi=0^{\circ}$, or $C_{2}>0$. Next, when $\Delta\sim\lambda_{R}$, we observe a Lifshitz transition, in Fig.~\ref{fig:rashba2b}(c). As in FeRh, the AMR corresponding to the onset of the Lifshitz transition contains multiple harmonics of comparable magnitude. Beyond the Lifshitz transition ($\Delta \ll \lambda_R$), the ferromagnetic Rashba model has two concentric Fermi surfaces when $|\vec{\Delta}|\rightarrow0$, which become distorted for non-zero $|\Delta|$, as seen in Fig.~\ref{fig:rashba2b}(d) (See Supplementary Note 3 for more details).

Nevertheless, it is clear that the change in the resistance of the $2$-band ferromagnetic Rashba model, computed as a function of order parameter strength and orientation, captures the essential features of the more complicated $8$-band model of FeRh. In particular, for small $|\vec{\Delta}|$ the AMR is predominantly two-fold symmetric, with a minimum of the resistance at $\phi=0$ ($C_{2}<0$) as in the extreme spin-orbit coupled limit of FeRh.

\subsection{AMR from Pseudogravitational Fields}
\label{sec:gravity2}

Both the more accurate $8$-band model of FeRh and the more simplified $2$ band model of the ferromagnetic Rashba metal point to the onset of a sign change in the AMR as a ubiquitous feature present in the AMR of spin-orbit coupled magnetic materials. Beginning with this observation, we propose a geometric framework within which to understand these effects that is rooted in the observation that the distortions of the Fermi surface in both the FeRh and ferromagnetic Rashba model bear a striking resemblance to the coupling of fermions to curved space. For a quadratically dispersing band, as in the ferromagnetic Rashba model with a small Fermi surface, the Fermi surface distortion may be parametrized by an effective low-energy Hamiltonian as
\begin{equation}
    H^{quad}_{eff} = g^{ij}(\Delta) k_{i} k_{j},
\label{eq:quadmetric}
\end{equation}
where the effective nontrelativistic ``metric'' tensor $g^{ij}(\Delta)$ is a function of the magnetic order parameter and its form determines the observed Fermi surface geometry. The geometric coupling arises due to the non-trivial SOC-induced winding of the spin texture on the Fermi surface interacting with the background magnetic order parameter, and therefore, is intimately connected band topology. For the Rashba model with a constant background magnetic order parameter, the metric describes an elliptical distortion of the Fermi surface; for larger Fermi surfaces, where quartic corrections to the dispersion become important, we recover Fermi surface geometries such as in Fig.~\ref{fig:rashba2b}.

 Due to the gravitational parallel we have exploited in our Fermi surface parametrization, where the coupling of the order parameter to the carriers results in a distorted Fermi surface, we refer to the coupling of the order parameter to the Fermi surface geometry as "pseudogravity". As we will see via the example of the Rashba model below, pseudogravitational fields are expected to arise whenever there is magnetic order in the presence of spin-orbit coupling. For systems such as FeRh which feature linearly-dispersing Weyl fermions in addition to quadratically dispersing Fermi pockets, we expect the Weyl fermions to be coupled to a relativistic pseudogravitational metric of the form~\cite{nissinen2019elasticity,landsteiner2011gravitational}
\begin{equation}
    g^{\mu\nu} = e^{\mu}_{\hphantom{\mu}\alpha}e^{\nu}_{\hphantom{\nu}\beta}\eta_{\alpha\beta} = \left(\begin{array}{c|c}
    -1 & -\mathbf{u} \\
    \hline
    -\mathbf{u} & AA^{T}
    \end{array}\right).\label{eq:metric}
\end{equation}
In Eq.~(\ref{eq:metric}), we show the metric in block matrix form expressed with the Minkowski signature typical of flat spacetime. The off-diagonal components of Eq.~(\ref{eq:metric}), $\mathbf{u}$ may be interpreted as the velocity of a moving frame, or a tilt in the case of a Weyl semimetal\cite{bradlyn2015low,nissinen2019elasticity}, and $A$ parametrizes the elliptical distortion of the Fermi surface.

Using the Rashba model as a simple example, we will now see how anisotropic magnetoresistance is a signature of pseudogravitation couplings. For a noninteracting system with nonmagnetic disorder, contributions to the ohmic conductivity originate from electrons near the Fermi surfaces, which, as in Fig.~\ref{fig:rashba2b}, distort as a function of $|\vec{\Delta}|$ and $\phi$. In particular, within the relaxation time approximation and with nonmagnetic disorder the dissipative conductivity depends on the geometry of bands near the Fermi surface, and is given by the Fermi surface integral
\begin{equation}
\sigma_{ij} \propto \int d^{2}k \left(\frac{\partial\epsilon_\mathbf{k}}{\partial k_{i}}\frac{\partial\epsilon_\mathbf{k}}{\partial k_{j}}\right)\delta(\mu-\epsilon_\mathbf{k}).\label{eq:cond}
\end{equation}
around the Fermi surface with energy $\epsilon_\mathbf{k}$. 
We separate the our analysis in the two regimes in the AMR that are of interest: $\Delta \gg \lambda_{R}$, where the $C_{2}$ harmonics dominate, and $\Delta \ll \lambda_{R}$, where similar $C_{2}$ harmonics dominate but with a $\frac{\pi}{2}$ phase shift. 

We begin by examining the limit $\Delta\gg \lambda$ where we seek both an expression for the effective metric and the conductivity. In this limit, we further restrict ourselves to negative values of the chemical potential $\mu$, such that we have a single Fermi surface. 

In this limit, we may write the Bloch Hamiltonian corresponding to Eq.~(\eqref{eq:rashbaham}) as
\begin{align}
H(\mathbf{k}) &= \sigma_0 [2t(2-\cos k_{x} - \cos k_{y})-\mu] +\Delta\cdot\vec{\sigma} + \lambda_{R}(\sin k_{x} \sigma_{y} -\sin k_{y}\sigma_{x})
\end{align}
Since, we have chosen $\mu$ such that the Fermi surface is small, we may expand the Hamiltonian about $\mathbf{k}=0$ to find
\begin{equation}
H(\mathbf{k})\approx [t(k_{x}^2+k_{y}^{2})-\mu]\sigma_0 + \Delta\cdot\vec{\sigma}+ \lambda_{R}(k_{x}\sigma_{y}-k_{y}\sigma_{x})
\end{equation}
with corresponding energies
\begin{equation}
E_{\pm} = t|k|^{2}-\mu \pm \sqrt{(\lambda_{R} k_{x} +\Delta\sin\phi)^{2}+(\lambda_{R} k_{y}-\Delta\cos\phi)^{2}} 
\label{eq:rashba_es}
\end{equation}
Let us now suppose that when $\Delta\gg \lambda_{R}$, the Fermi surface lies entirely in the $E_{-}$ band. By expanding occupied band, $E_{-}$, in powers of $\lambda_{R}/\Delta$ to find the effective Hamiltonian that is quadratic in momentum, as expected from Eq.\eqref{eq:quadmetric}, or
\begin{equation}
E_{-} \approx \Delta + t|k|^{2}-\mu + \lambda_{R}(k_{x}\sin\phi-k_{y}\cos\phi)+\frac{\lambda_{R}^{2}}{2\Delta}(k_{x}\cos\phi+k_{y}\sin\phi)^{2}.
\label{eq:elowband}
\end{equation}
We see that, to lowest order the energy, $E_{-}$ is a quadratic form in momentum, $\mathbf{k}$, and we may complete the square to write,
\begin{equation}
E_{-} = E_{-0} + g_{ij}(\phi)k_{i}'k_{j}',
\label{eq:elowband2}    
\end{equation}
where to lowest order we may make the following substitutions in Eq.(\eqref{eq:elowband}) to arrive at the Eq.(\eqref{eq:elowband2})
\begin{align}
k_{x}' &= k_{x} + \lambda_{R}\sin\phi/2t \\
k_{y}' &= k_{y}-\lambda_{R}\cos\phi/2t \\
E_{0-} &= \Delta-\mu + \lambda_{R}^{2}\begin{pmatrix}
\sin\phi& -\cos\phi
\end{pmatrix}_{i}g_{ij}\begin{pmatrix}
\sin\phi \\
-\cos\phi
\end{pmatrix}_{j}\\
g_{ij} &= \begin{pmatrix}
t+\frac{\lambda_{R}^{2}}{2\Delta}\cos^{2}\phi & \sin\phi\cos\phi\frac{\lambda_{R}^{2}}{2\Delta} \\
\sin\phi\cos\phi\frac{\lambda_{R}^{2}}{2\Delta} & t+\frac{\lambda_{R}^{2}}{2\Delta}\sin^{2}\phi
\end{pmatrix}_{ij}\label{eq:rashbametricconcrete}.
\end{align}
Eq.~\eqref{eq:rashbametricconcrete} gives the pseudogravitational metric $g_{ij}(\phi)$ as a function of the tight-binding parameters. The metric $g_{ij}(\phi)$ naturally emerges from the parametrization after the low-momentum expansion. The presence of the pseudogravitational metric is expected here as it determines the geometry of the Fermi surface which, in turn, sets the velocity at the Fermi surface.  Substituting Eq.~(\eqref{eq:elowband2}) this into Eq.~(\eqref{eq:cond}) for the conductivity, we find that
\begin{equation}
\sigma_{ij}(\phi)) = \frac{1}{\sqrt{\det g_{ij}(|\Delta|,\phi)}}g_{ij}(|\Delta|,\phi)\bar{\sigma_0}.
\label{eq:siggy}
\end{equation}
where $\bar{\sigma_0}$ is a $\phi$-independent but $\mu$ and $t$-dependent constant. Eq.~\eqref{eq:siggy} shows that the conductivity tensor in this limit is a function of the pseudogravitational metric. To find the AMR, we invert Eq.~(\eqref{eq:siggy}) resulting in an equation for the longitudinal resistivity as a function of $\phi$, or
\begin{equation}
\rho_{xx}(\phi) = \bar{\rho} + \frac{\lambda^{2}}{4t\Delta}\cos 2\phi + \mathcal{O}(\lambda_{R}^{3}).
\label{eq:delta-amr}
\end{equation}
Eq.~(\eqref{eq:delta-amr}) reproduces the positive two-fold dominant, $C_{2}$, AMR observed in the high-field magnetic field limit of both the $8$-band tight-binding model for FeRh and the $2$-band Rashba model. In our analysis, we note that including the next-order correction to Eq.~\eqref{eq:delta-amr} produces a term that is proportional to $\cos4\phi$. The $C_{4}$ dominant Fourier harmonic is observed both experimentally and numerically when $\Delta \approx \lambda_{R}$ when the Fermi surface is close to the onset of the Lifshitz transition. 

Next, we examine the conductivity in the low magnetic field limit, or when $\Delta \ll \lambda_{R}$. In this limit, we expand the energies $E_{\pm}$ in Eq.~(\eqref{eq:rashba_es}) in powers of $\Delta/\lambda_{R}\ll 1$. We take advantage of the fact that the spin-orbit coupling, $\lambda_{R}$, multiplies the momentum, $|\vec{k}|$, in Eq.~\eqref{eq:rashba_es} to make a change of variables such that
\begin{equation}
\vec{k'} = \vec{k} + \frac{\Delta}{\lambda_{R}}(\sin \phi,-\cos \phi).
\end{equation}
In terms of adjusted momentum, $\vec{k'}$, we may rewrite the energies for the $2$-band Hamiltonian of Eq.~\eqref{eq:rashba_es} in the low-field limit as

\begin{equation}
E_\pm= t|\vec{k}'|^{2} \pm \lambda_{R}|\vec{k}'|\left(1 \mp \frac{2t\Delta}{\lambda_{R}^{2}}\sin(\phi-\theta)\right)+\frac{t\Delta^{2}}{\lambda_{R}^{2}}-\mu
\label{eq:rashba_e_large_ld}
\end{equation}

We see that, distinct from the $\Delta \gg \lambda_R$ limit, in the $\Delta\ll\lambda_R$ limit the constant $E_\pm$ contours are anisotropic and depend explicitly on the angle $\theta-\phi$ between $\vec{k'}$ and $\vec{\Delta}$. This dependence, combined with the explicit dependence of the energy on $|\vec{k}'|$ indicates that the constant energy surfaces of Eq.~\ref{eq:rashba_e_large_ld} do not admit a description in terms of a quadratic form; in this limit, the Fermi surfaces of the Rashba model are quartic surfaces. Nevertheless, we observe that by rotating the order parameter, or changing $\phi$ induces a distortion of the constant energy contours of Eq.~\eqref{eq:rashba_e_large_ld}. 

In a similar fashion to the high-field limit, we may use Eq.~\eqref{eq:cond} to evaluate the conductivity of the Rashba model perturbatively in $\Delta/\lambda_{R}$. In the low-field limit, however, we find that the anisotropy in the conductivity tensor arises due to the $\vec{\Delta}$ dependence of the quartic Fermi surface shape and, by extension, the Fermi velocity. The presence of the quartic Fermi surface in this limit does not admit a simple analytic solution, as was the case in the high-field limit. On the other hand, in the limit that the chemical potential, $\mu$ lies in between the upper and lower spin-split bands, ensuring that there is only a single large Fermi surface, we may numerically solve for the conductivity and resistivity to find that the AMR for the Rashba model is $C_{2}$ dominant and negative in the $\Delta \ll \lambda_{R}$ limit.

We may now understand the core physical rationale behind the appearance of the Lifshitz transition in the FeRh and ferromagnetic Rashba AMR. In both the $\Delta\ll\lambda_R$ and $\Delta\gg\lambda_R$ limits, we have a strong interdependence between the magnetic order parameter and the spin-orbit coupling. Therefore, under the application of a strong in-plane magnetic field, the pseudogravitational fields couple to the electrons on the Fermi surface leading to the distortions, seen in Figs.~(\ref{fig:LDOS3P}) and (\ref{fig:rashba2b}). Continued increases in magnetic field leads to increasing pseudogravitational modification of the Fermi surface that pulls the connected Fermi pockets apart leading to the observed features in the AMR. As we see in Eq.~\eqref{eq:siggy}, when the Fermi surface is approximately quadratic, the pseudogravitational distortion directly determines the conductivity tensor and hence the AMR.

The end result is that we arrive at a simple interpretation of the observed AMR: the interplay between spin-orbit coupling and magnetic order creates a pseudogravitational coupling between the order parameter, the Fermi surface shape that alters the resultant quasiparticle velocity. The AMR in our $8$-band model of FeRh arises from a similar mechanism and, although the coupling between the order parameter and the $4$ bands crossing the Fermi surface is more intricate in FeRh, we nevertheless observe in Fig.~\ref{fig:rashba2b} that the AFM order parameter distorts the Fermi surfaces leading to the observed AMR. We note that our calculations for both the Rashba and $8$-band FeRh model treat disorder as non-magnetic in nature within the relaxation time approximation. However, since the corresponding self-energies from relevant phonon or spin-fluctuation contributions are small in magnitude, their presence would not impact the results presented in this work\cite{YKMJGPhysRevB98}. 

\section{Conclusions}
\label{sec:Outlook}

In this work, we have measured the anisotropic magnetoresistance of thin-film FeRh samples in the antiferromagnetic regime. We showed that as the magnetic field rotates, the order parameter smoothly tracks the magnetic field direction. Through a combination of ab-initio calculations and tight-binding modeling of thin-film FeRh, we showed how the observed AMR is a result of the evolution of the Fermi surface geometry as a function of applied magnetic field. We demonstrated that the coupling between the inherent magnetic order and Fermi surface geometry is ubiquitous in spin-orbit coupled magnets and is responsible for the most salient observables in the AMR measurements. Using a simplified ferromagnetic Rashba model, we are able to illustrate the origin of this coupling: spin-orbit coupling induces Fermi-surface spin textures that are influenced by the magnetic order. With the aid of a $\mathbf{k}\cdot\mathbf{p}$ approximation for small Fermi surfaces in the Rashba model, we find that the distortion of the Fermi surface plays the role of an anisotropic band mass tensor that depends on the magnetic order parameter. From a theoretical perspective, this tunable band mass anisotropy is analogous to an effective pseudogravitational metric,
\begin{equation}
\frac{1}{2m}\sum_{i} k_{i}^{2} \rightarrow \sum_{ij}g^{ij}(|\Delta|,\phi) k_{i} k_{j},
\end{equation}
that allowed for a much deeper understanding of the underlying physics associated with the Lifshitz transition. In FeRh, we have linearly dispersing Weyl pockets near the Fermi level in addition to typical quadratic metallic Fermi surfaces. As such, we expect these Weyl Fermi surfaces are also similarly distorted due to the coupling between spin-orbit interaction and magnetism. Using the analysis associated with the pseudogravitational mapping will shed light on observations in other magnetic systems that possess both topological and mean-field order such as EuTiO$_{3}$\cite{ahadi2019anisotropic}. In a broader context, our results suggest that easy-plane antiferromagnets with strong spin-orbit coupling are candidate systems for exploring the coupling of fermions to distorted background geometries. 
 
\section*{Acknowledgements}
This research was primarily supported by the NSF through the University of Illinois at Urbana-Champaign Materials Research Science and Engineering Center DMR-1720633. Thin-film growth at Argonne National Laboratory was supported by the U.S. Department of Energy, Office of Science, Materials Science and Engineering Division. B.B. acknowledges the support of the Alfred P. Sloan foundation, and the National Science Foundation under grant DMR-1945058. M.G. and J.O. acknowledge support from ONR 14-17-1-3012. M.G.V. thanks support from the Spanish Ministry of Science and Innovation (grant number PID2019-109905GB-C21) and thanks F. Orlandi for discussions.  

\section*{Author Contributions}
J.S and N.M. designed the magnetotransport experiments. H.S. and A.H. were responsible for synthesizing FeRh films. J.S., S.S, and H.S. performed magnetic and structural characterization of FeRh. J.S, J.O., and H.S. fabricated devices. J.S, S.S, and J.O. performed magnetotransport experiments. A.H. and N.M. supervised the experimental work. M.V. performed DFT calculations with input from M.G. and B.B. on magnetic structure. M.G. and B.B. developed the theoretical models and quantum transport simulations. M.G. and B.B. analyzed the results with J.S., A.H. and N.M. providing input. B.B. and M.G. wrote the manuscript with input from all other authors. 

\section*{Competing Interests}
The authors declare no competing interests.

\bibliographystyle{unsrt}
\section{Methods}

\subsection{$FeRh$ Crystal Growth}
The FeRh films used in this work have been deposited onto $[001]$-oriented MgO substrates using DC magnetron sputtering. Prior to sputtering, the substrates are heated within the sputter deposition system to $850^\circ ~ C$ for one hour to ensure that potential contaminants are desorbed from the surface.  After the substrates are cleaned, the temperature was lowered to 450$^\circ$ C for deposition.  The sputter target used for deposition is an equiatomic FeRh source.  During growth, $6.5$ sccm of Ar gas was introduced into the chamber, and the pressure was set to $6$~mTorr.  The DC sputtering power used was $50$ W, and the growth rate was $0.7 \mathrm{\AA}$/s. 

\subsection{$FeRh$ Device Fabrication}
A photolithographically patterned mask has been developed onto a continuous film in the intended Hall bar pattern, and an ion-milling process removes the FeRh film not under the mask.  Magnetotransport measurements were made inside a Quantum Design, Physical Properties Measurement System (PPMS).  Longitudinal magnetoresistance and transverse voltage measurements of samples were made using standard lock in detection with a Stanford Research Systems SR830 lock in amplifier.  To facilitate lock-in detection, a $17$ Hz probe current of nominally 10~$\mu$A was used.

\subsection{Density Functional Calculations of $FeRh$}
To perform DFT calculations, we use the Vienna Ab-Initio Simulation Package \cite{VASP1,VASP2} (VASP). The projected augmented wave (PAW) pseudopotentials\cite{vaspPaw} are used for the calculations and the exchange-correlation energy is calculated with generalized gradient approximation (GGA) in the Perdew-Burke-Ernzerof form\cite{PBE}. A $\Gamma$-centered 11 $\times$ 11 $\times$ 11 Monkhorst-Pack $k$-mesh is used for the calculations. We reproduce the Dzyaloshinskii-Moriya interaction (DMI) using the constrained moment approach with spin-orbit coupling interactions implemented in VASP so as to calculate the band structure for different spin orientations \cite{DMI1,DMI2,DMI3}.  In order to reproduce the strong correlation of the $d$-orbitals that drives the magnetism, we use a Hubbard $U$ of $3$~eV for both magnetic atoms. In order to understand the evolution of the bandstructure under the experimental conditions presented in the main text, we must be able to approximate the effects of an in-plane magnetic field on the thin-film FeRh. In our work, we mimic the magnetic field by configuring the normally non-magnetic Rh magnetic moments to be configured in such a manner so as to possess an in-plane ferromagnetic orientation. One of the important consequences of the Rh atoms developing an in-plane ferromagnetic state is that the nature of the FeRh crystal changes from symmorphic to non-symmorphic. 
\section*{Data Availabilty}
The data that support the findings of this study are available from the corresponding author upon reasonable request.
\newpage

\vspace{50mm}
\section*{References}
\bibliography{Weyl}

\end{document}


\title{Supplemental Material: Evidence of Pseudogravitational Fields in Spin-Orbit Coupled Antiferromagnet FeRh }

\author{Joseph Sklenar}
\affiliation{Department of Physics and Astronomy, Wayne State University, Detroit, MI 48201, USA}
\affiliation{Department of Physics, University of Illinois at Urbana-Champaign, Urbana, IL 61801, USA}
\affiliation{Materials Research Laboratory, University of Illinois at Urbana-Champaign, Urbana, IL 61801, USA}
\author{Soho Shim}
\affiliation{Department of Physics, University of Illinois at Urbana-Champaign, Urbana, IL 61801, USA}
\affiliation{Materials Research Laboratory, University of Illinois at Urbana-Champaign, Urbana, IL 61801, USA}
\author{Hilal Saglam}
\affiliation{Materials Science Division, Argonne National Laboratory, Lemont IL 60439, USA}
\affiliation{Department of Physics, Yale University, New Haven, Connecticut 06520, USA}
\author{Junseok Oh}
\affiliation{Department of Physics, University of Illinois at Urbana-Champaign, Urbana, IL 61801, USA}
\affiliation{Materials Research Laboratory, University of Illinois at Urbana-Champaign, Urbana, IL 61801, USA}
\author{M. G. Vergniory}
\affiliation{Donostia International Physics Center, P. 
Manuel de Lardizabal 4, 20018 Donostia–San Sebastian, Spain}
\affiliation{Max Planck Institute for Chemical Physics of Solids, Dresden, D-01187, Germany.}
\author{Axel Hoffmann}%
\affiliation{Materials Science Division, Argonne National Laboratory, Lemont IL 60439, USA}
\affiliation{ Department of Materials Science and Engineering, University of Illinois at Urbana-Champaign, Urbana, IL 61801, USA
}%
\affiliation{ Department of Electrical and Computer Engineering, University of Illinois at Urbana-Champaign, Urbana, IL 61801, USA
}%
\affiliation{Materials Research Laboratory, University of Illinois at Urbana-Champaign, Urbana, IL 61801, USA}
\affiliation{Department of Physics, University of Illinois at Urbana-Champaign, Urbana, IL 61801, USA}
\author{Barry Bradlyn}
\affiliation{Department of Physics, University of Illinois at Urbana-Champaign, Urbana, IL 61801, USA}
\affiliation{Institute for Condensed Matter Theory, University of Illinois at Urbana-Champaign, Urbana, IL 61801, USA}
\author{Nadya Mason}
\affiliation{Department of Physics, University of Illinois at Urbana-Champaign, Urbana, IL 61801, USA}
\affiliation{Materials Research Laboratory, University of Illinois at Urbana-Champaign, Urbana, IL 61801, USA}
\author{Matthew J. 
Gilbert}
\affiliation{ 
Department of Electrical and Computer Engineering, University of Illinois at Urbana-Champaign,  Urbana, IL 61801, USA
}%

\maketitle


\date{\today}

\section*{Supplementary Note 1: Additional Experimental Characterization}

\subsection{X-ray Diffraction Measurements}
\label{sec:sup-xrd}
\begin{figure*}
\includegraphics[width=\textwidth]{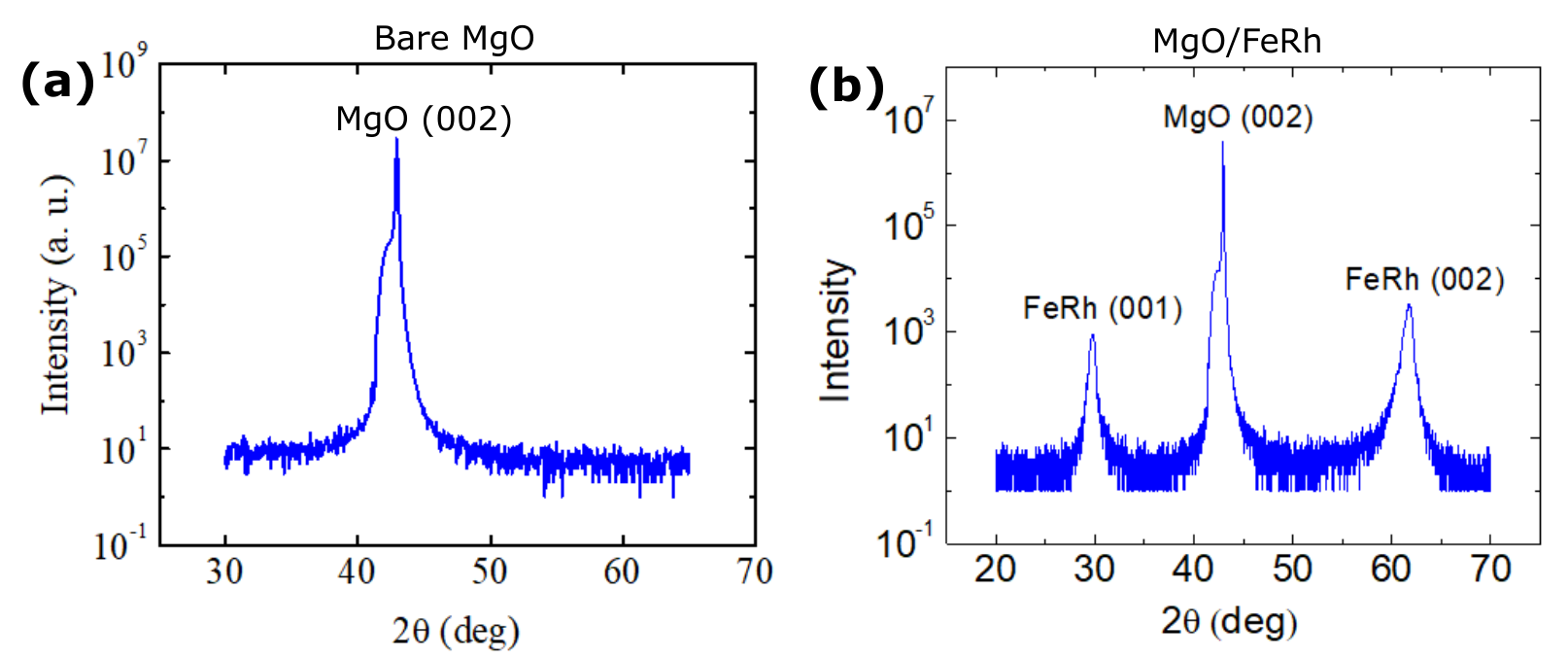}
\caption{\textbf{X-ray Diffraction of FeRh:} (a) XRD of a bare, [001]-oriented, MgO substrate.  (b)  XRD of a $20$ nm thick FeRh film deposited onto an MgO substrate}
\label{fig:XRD}
\end{figure*}

In Supplemental Figure~\ref{fig:XRD} (a) and (b), X-ray diffraction (XRD) measurements are shown for both a $[001]$-oriented MgO substrate along with the MgO substrate with a FeRh films deposited upon the surface.  A Bragg peak corresponding to the $(002)$ family of lattice planes in MgO is present for both measurements. In (b) that there are two additional peaks corresponding to the $(001)$ and $(002)$ family of planes due to the FeRh film.  The peak positions correspond to lattice constants along the $c$-axis of $0.421$ nm and $0.300$ nm for MgO and FeRh respectively.  Because the $[100]$ orientation of FeRh tends to grow along the $[110]$ direction of the MgO substrate, these measurements correspond to a lattice mismatch of approximately $0.3\%$. These parameters are consistent with other reports in the literature, and are indicative of epitaxial growth of FeRh on the MgO.

\subsection{Magnetometry and the Metamagnetic Transition}
\label{sec:sup-metamag}
In FeRh, antiferromagnetism is present at low temperatures after the sample has undergone the metamagnetic transition.  There are two complementary ways to verify the metamagnetic transition in our FeRh samples.  Measuring the magnetic moment of FeRh films with superconducting quantum interference (SQUID) magnetometry is a commonly used method to verify the transition exists. Experimentally, one first applies an external field in the high-temperature ferromagnetic phase.  Next the temperature is lowered, and at the onset of the ferro- to antiferromagnetic transition a sudden drop in the magnetization will occur.  As the temperature continues to decrease, no further abrupt changes in the magnetization appear.  Upon warming up, with the same applied field present, a sudden increase in the magnetization will occur at the onset of the antiferro- to ferromagnetic transition.  Because the metamagnetic transition is a first-order phase transition, the transition temperatures exhibit hysteresis depending on the initial magnetic state of the FeRh.  In Supplemental Figure ~\ref{fig:Metamag} (a) the metamagnetic transition for the $20$ nm thick film is shown with a $0.2$ T field applied in-plane.  As described in the main text, thin FeRh films have a ferromagnetic region near the MgO interface due to strain, and this ferromagnetism persists even at low temperatures.  This residual ferromagnetic moment is most easily observed by measuring the magnetization of the sample as a function of magnetic field both above and below the metamagnetic transition.  In Supplemental Figure~\ref{fig:Metamag} (b), we show two representative magnetic hysteresis loops as the field is swept at $350$ K and at $100$ K, in the ferro- and antiferromagnetic phase respectively.  The residual moment in the antiferromagnetic phase at 100 K  is roughly $20\%$ of the saturated moment in the ferromagnetic phase at $350$ K.

The metamagnetic transition may be detected electrically by measuring the resistance of a sample as a function of temperature.  It is well-established that an increase in resistivity accompanies the transition from ferro- to antiferromagnetic order.  In Supplemental Figure ~\ref{fig:Metamag} (c), we plot the resistance versus temperature for the Hall bar sample with current along the $[100]$ direction of FeRh.  As indicated by the arrows, both a cool-down and warm-up measurement are present in the data.  In this representative trace, a $7$ T field is for both the cool-down and warm-up measurement sequence.  By comparing the metamagnetic transition observed with SQUID magnetometry in (a), versus electrical resistivity in (c), we see the influence that the external field has on the transition.  Namely, when a $0.2$ T field is present the transition is just below $300$ K, and when a $7$ T field is present the transition is closer to $230$ K. 
\begin{figure*}
\includegraphics[width=\textwidth]{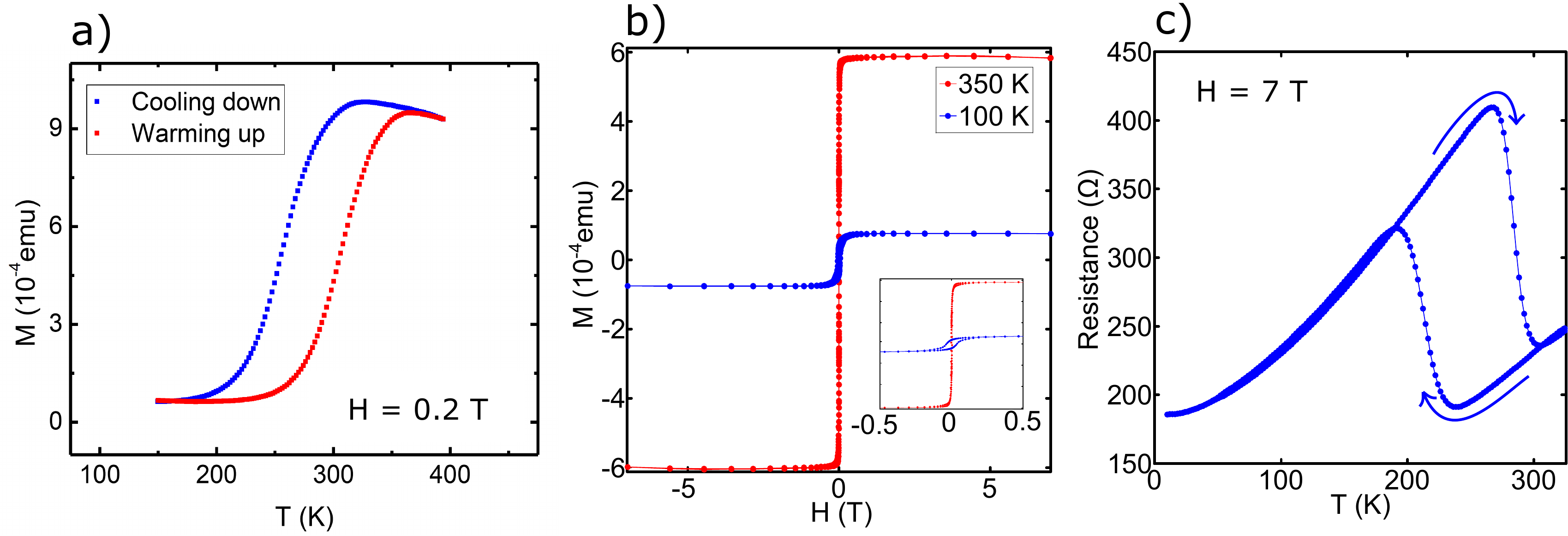}
\caption{\textbf{Magnetometry Measurements on $FeRh$:}(a) The magnetization of a $20$ nm thick FeRh film is measured as a function of temperature in the presence of a $0.2$ T external magnetic field.  The blue curve indicates that the temperature is decreasing while the red curve indicates the temperature is increasing. (b)  The magnetization as a function of field is measured in the ferromagnetic phase (red curve) at $350$ K, and in the antiferromagnetic phase (blue curve) at $100$ K.  Note that some residual ferromagnetic moment remains in the antiferromagnetic phase.  As shown in the inset, the residual ferromagnetism within the film has a coercive field well below $500$ mT.  (c)  The resistance of the same FeRh Hall bar, where the AMR was presented in Fig.~$2$ of the main text, is measured as a function of temperature.  Here, a $7$ T magnetic field is fixed as the temperature is lowered from $310$ K to $2$ K, and then increased back to $310$ K.  The blue arrows indicate the direction of the temperature sweep for a given branch of the curves.  There is clear hysteresis in the resistivity indicating the metamagnetic transition.  The upper branch indicates the antiferromagnetic phase and the lower branch indicates the ferromagnetic phase.  Note that for (a), (b), and (c), the external field is within the plane of the film.  }
\label{fig:Metamag}
\end{figure*}
%
\section*{Supplementary Note 2: $FeRh$ Anisotropic Magnetoresistance: Current in the $[110]$ Crystal Direction}
\label{sec:110amr}

In this section, we examine the AMR when the current is applied along the $[110]$ direction of FeRh. In Supplemental Figure~\ref{fig:spectral110}(a), we show the experimental angular dependence of the AMR for in-plane magnetic fields ranging from $1~T$ to $12~T$. We observe a dominant four-fold symmetric contribution that is due to different initial MCA-dependent magnetization orientations present after the field cooling cycle. The four-fold dominant AMR and the reduced AMR magnitude indicates that thin-film FeRh has an increased magnetic susceptibility along the $[110]$ crystal direction. As a result, the spin-orbit coupling, $\lambda(\phi,B)$ depends on both the current direction and the external magnetic field due to details of the sample preparation. 

Furthermore, the use of one static isotropic value of spin-orbit interaction in the $(x-y)$ plane neglects the presence of magneto-crystalline anisotropy (MCA). To accommodate for the presence of MCA consistent with the $d$-orbital symmetry in the $(x-y)$ plane, we modify the value of $\lambda$ to be $\lambda=\lambda_{xy}\times\left(1-0.2\cos(4\varphi)\right)$, where $\varphi$ is the angle of deviation between the crystal direction of the current flow direction and the $[100]$ crystal direction. 

To understand the transport dynamics, we apply the modified tight-binding model to calculate the AMR when current is in the $[110]$ direction. Similar to our analysis in Sec.~$III~B$ of the main text, the antiferromagnetic exchange, cant angle, and ferromagnetic Rh moment in the $8$-band model are assumed to follow identical linear trends as the magnetic field increases in the experimental measurements.

We show the results of the calculation in Supplemental Figure~\ref{fig:spectral110}(b). We that our model accurately captures the essential features of the experimental AMR:  a dominant four-fold harmonic and small changes in the magnitude of the AMR over the range of cant angles we consider. As before, we perform a spectral decomposition of the AMR signals in both the theory and experiment. Beginning with the experimental AMR, shown in Supplemental Figure~\ref{fig:spectral110}(c), we find that for low cant angles, the AMR is twofold dominant with $C_{2} > C_{4}$. There is a crossover $\theta^{even}_{cant}\approx 3.0^{\circ}$ after which $C_{4}$ is dominant for all larger field strengths we examine. Interestingly, $C_{2}$ contribution changes sign as the canting angle increases with the crossover occurring at $\theta^{C_{2}}_{cant} \approx 15.0^{\circ}$. This is similar to the behavior of $C_{2}$ in the $[100]$ AMR measurements, pointing towards the universality of this effect. Note that, unlike in our $[100]$ measurements, the sign change in $C_{2}$ is not as visibly apparent due to the size of $C_{4}$. Comparing to the spectral decomposition of the theoretical AMR, in Supplemental Figure~\ref{fig:spectral110}(d), we confirm the qualitative similarities observed when comparing the AMR in the $[110]$ direction. 

We find that the theoretical calculations reproduce the correct trend in $C_{2}$. For low values of $\theta_{cant}$ both the theoretical AMR and the experimental measurements have $C_{2}\sim C_{4}$. As the canting angle increases, we continue to observe a theoretical AMR dominated by $C_{4}$. 
The theoretically calculated AMR shows a sign reversal of $C_{2}$, at a cant angle of approximately $7^\circ$, which is smaller than the experimentally derived value for the transition. As $\theta_{cant}$ increases past $7^\circ$, the theoretical model additionally predicts a crossover with $C_2 > C_4$ that is not observed in the experimental data at the magnetic fields considered.

\begin{figure*}
\begin{centering}
\includegraphics[width=\textwidth]{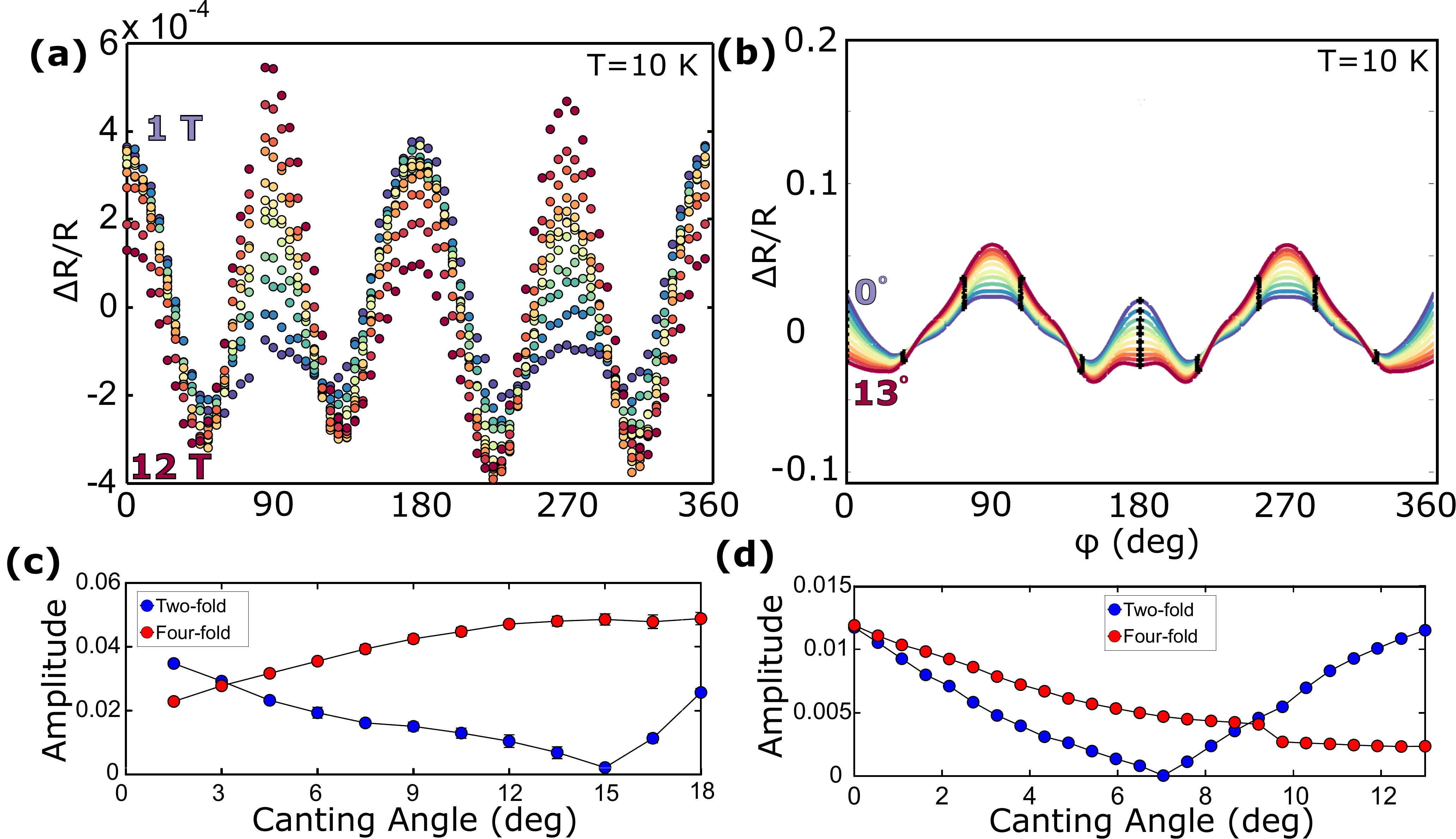}
\caption{\textbf{Quantum Transport in the $[110]$ Crystal Direction:} (a) Experimentally measured AMR for in-plane magnetic fields between $1$-$12$ $T$. In contrast to the AMR measurements in the $[100]$ direction, the magnitude of the AMR is suppressed and the harmonics are dominated by $C_{4}$ symmetric components. (b) Theoretically calculated AMR for canting angles, $\theta_{cant}$ between the Fe moments in the range of $0^\circ$-$13^\circ$. The reduction in the cant angle range arises from the increased spin-orbit interactions present in the $[110]$ as compared to the $[100]$. The spectral decomposition of the experimental and the theoretical AMR amplitudes are shown in (c) and (d) respectively. We observe qualitative agreement between the theoretical and experimental spectral components in that the most prominent features inherent in the harmonic analysis, such as the change in sign of the $C_{2}$ symmetric component, are present in both the experimental and theoretical AMR}
\label{fig:spectral110}
\end{centering}
\end{figure*}

\subsection{Comparison with the $[100]$ Data}

The AMR with current along $[110]$ provides a starkly different picture in terms of the spectral contribution and the overall magnetic susceptibility of FeRh observed in the $[100]$ direction. We attribute these differences to the in-plane MCA in FeRh, which manifests as a decreased magnetic susceptibility when the in-plane magnetic field is increased in magnitude along the $[110]$ direction as compared to the $[100]$ direction. This is consistent with previous studies showing similar magnetization insensitivity under application of large in-plane magnetic fields\cite{marti2014room} albeit in thicker films where: (1) the effects of the ferromagnetic underlayer of FeRh that appears at the interface and (2) the magnetization near the MgO interface resulting from magnetic field polarization are mitigated. In our thin-films, the ferromagnetic interactions from substrate and interface effects successfully compete with the antiferromagnetic exchange thereby allowing for greater magnetization dynamics. Thus, although we expect the order parameter to remain parallel to the applied magnetic field in both the $[100]$ and $[110]$ measurements, due to the small thickness of our samples there is reduced magnetic susceptibility in the $[110]$ measurement. The reduced magnetic susceptibility leads to smaller cant angles, and, therefore, a weaker dependence of the Hamiltonian on the external field.

\section*{Supplementary Note 3: $FeRh$ Theoretical Details}

\subsection{Density Functional Theory Details}
\label{sec:dft}

Experimental results, presented in Sec.~\ref{sec:sup-xrd}, indicate that after the FeRh is deposited onto the MgO substrate, biaxial strain from lattice mismatch causes a lattice distortion that alters the normal cubic configuration of FeRh to a tetragonal or an orthorombic configuration, as described in Ref.~\cite{zarkevich2018ferh}. The AFM configuration resulting from the induced distortion is depicted in Supplemental Figure \ref{fig:XS} and is described by the $P_{b}2/m$ (No. 10.48) magnetic group. The $P_{b}2/m$ space group has a monoclinic unit cell that is two times larger than the primitive cell of the nonmagnetic configuration. In the calculations presented here, we utilize the lattice constants and atomic coordinates reported in Ref. \cite{zarkevich2018ferh} in order to be consistent with previous known work on FeRh. 
\begin{figure}
 \begin{center}
 \label{fig:dappl1}
 \includegraphics[width=3.25in,angle=-90]{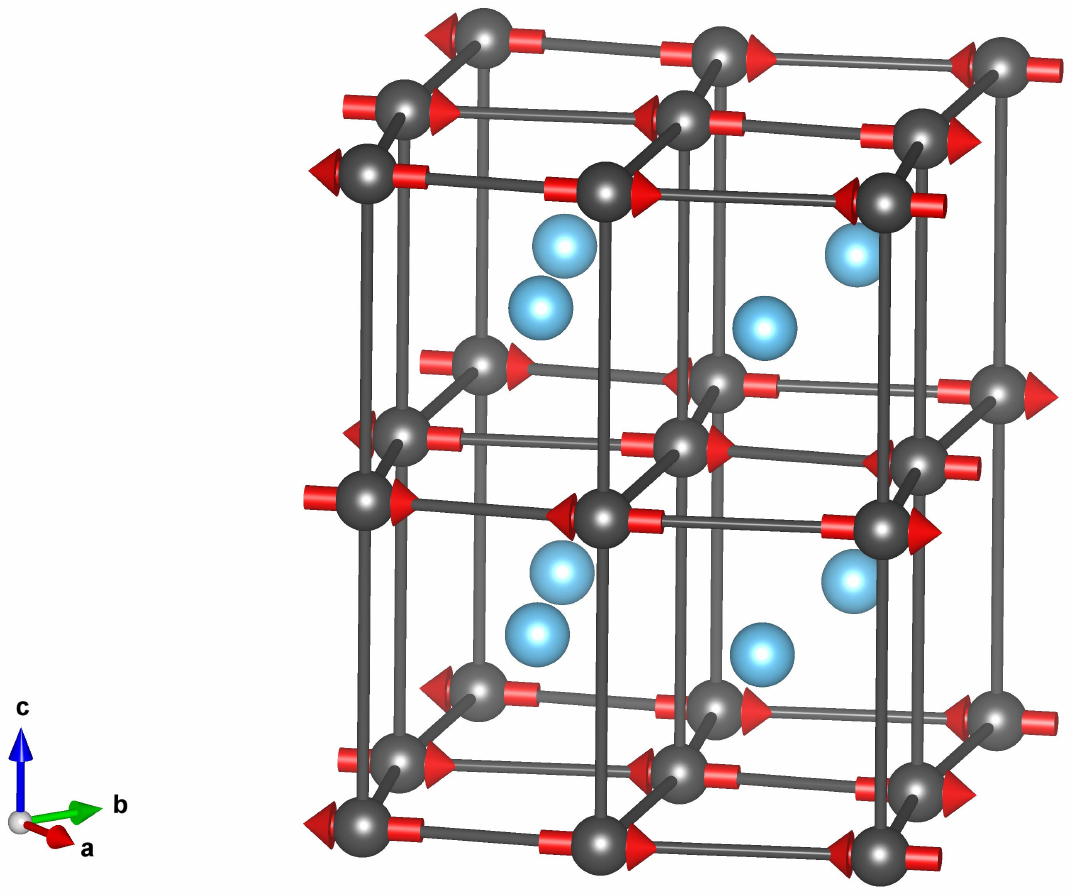}
\caption{\textbf{Lattice representation of the orthorhombic configuration of the antiferromagnetic phase of FeRh:} In this representation, the Fe atoms, shown in grey, possess the magnetic moments, and the Rh atoms, shown in blue, do not possess any magnetism. Therefore, in the absence of an in-plane magnetic field, FeRh is a symmorphic crystal.  }
\label{fig:XS}
\end{center}
\end{figure}
%
\subsection{Tight-Binding Model for FeRh}
\label{sec:sup-tbdef}

In order to understand the AMR, we construct a tight-binding model using the general Hamiltonian
\begin{equation}
\begin{split}
 \mathrm{H}  = \sum\limits_{\langle \langle i,j \rangle \rangle} t_{ij} c^{\dagger}_{i} c_{j} + \sum \limits_{\langle \langle i,j \rangle \rangle, {\langle l \rangle}} c^{\dagger}_{i} i\vec{\lambda_{ij}}  \cdot \Vec{\bf{\sigma}} c_{j} \\
 + \Delta \sum\limits_{i} \xi c^{\dagger}_{i} ( \Vec{{\bf{\sigma}}} \cdot \Vec{\bf{m}}) c_i
 \end{split}
 \label{eq:hamtbs}
\end{equation}

\subsubsection{Non-Magnetic Matrix Elements}

We begin the first term in Eq.~(\ref{eq:hamtbs}) splitting it into both intralayer (here defined to be hopping between atoms of a different species) and interlayer (hopping between atomic sites of the same species) nearest neighbor and next-nearest neighbor contributions\cite{wang2017antiferromagnetic,wang2017magnetic}. In order to define the $8$-band Hamiltonian, we make use of $4\times4$ affine rotation matrices. The first of these matrices are the lattice translation matrices for both left, $\Sigma_{L}$, and right, $\Sigma_{R}$ translations. We define the left translation matrix, $\Sigma_{L}$, as
\begin{equation}
\Sigma_{L} = \Sigma_{R}^{\dagger} =
\begin{bmatrix}
 0 & 0 & 0 & 1 \\
 1 & 0 & 0 & 0 \\
0 & 1 & 0 & 0 \\
0 & 0 & 1 & 0
\end{bmatrix}.
\end{equation}
Within the tight-binding formulation of thin-film FeRh, the matrices needed are variants of $\Sigma_{L}$ and $\Sigma_{R}$ that we define as
\begin{equation}
\Sigma_{L/R}^{0}=\Sigma_{L/R}\otimes\sigma_{0}.
\end{equation}
In the construction of the $4$ atom tight-binding model, we have, to this point, neglected the relative positions of the atoms. We begin with a layer of Fe atoms in the cubic configuration with the constituent Fe atom within the unit cell located at position $(0, 0, 0)$ in real-space. The next Fe atom within the same plane of Fe atoms are thus located at $(\pm a \hat{x}, \pm a \hat{y}, 0)$. where $a$ is the FeRh lattice constant. On top of the first layer of Fe atoms is a layer comprised entirely of Rh atoms that are translated by $(\pm \frac{a}{2} \hat{x}, \pm \frac{a}{2} \hat{y},0)$ away from the first Fe atom located at $(0, 0, 0)$. In this manner, the Rh atom that is included in the reduced unit cell is located at $(\frac{a}{2}\hat{x}, \frac{a}{2}\hat{y}, 0)$ The third layer of atoms, once again, consists of Fe atoms that are located within the layer in a cubic orientation. Relative to the first layer of Fe atoms, the third layer of atoms is located $(0, 0, \pm a\hat{z})$ with the member of the unit cell is located at $(0, 0, a\hat{z})$. In the fourth, and final, layer consists of Rh atoms that sit directly above the second layer of Rh atoms, once again, translated by $(\pm \frac{a}{2} \hat{x}, \pm \frac{a}{2} \hat{y}, \pm a \hat{z})$ from the Fe layers. Therefore, the final atom in the unit cell is a Rh atom located at $(\frac{a}{2} \hat{x}, \frac{a}{2} \hat{y}, \pm a \hat{z})$. Therefore, the locations of the $4$ atoms that form the unit cell for FeRh are: $Fe_{1} \rightarrow (0\hat{x}, 0\hat{y}, 0\hat{z})$, $Rh_{1} \rightarrow (\frac{a}{2}\hat{x}, \frac{a}{2}\hat{y}, 0\hat{z})$, $Fe_{2} \rightarrow (0\hat{x}, 0\hat{y}, a\hat{z})$ $Rh_{2} \rightarrow (\frac{a}{2}\hat{x}, \frac{a}{2}\hat{y}, a\hat{z})$.

We consider first the intralayer hopping from the Fe atom to the Rh atom independent of the spin at a distance of $(\frac{a}{2}, \frac{a}{2}, 0)$ away assuming that the initial Fe atom is located at $(0, 0, 0)$ in real-space. The nearest-neighbor intralayer contribution to the Hamiltonian is
\begin{equation}
H_{intra} = \left[t_{xy} \cos(\frac{k_{x}}{2})\cos(\frac{k_{y}}{2})\right] \cdot (\Sigma_{R}^{0} + \Sigma_{L}^{0}).
\label{eq:hintra_nn}
\end{equation}
In Eq.~(\ref{eq:hintra_nn}), $t_{xy}$ is the magnitude of the intralayer hopping. Using the same summations, we construct the interlayer hopping contribution to the FeRh Hamiltonian moving from an Fe atom to an Rh atom in the out-of-plane (interplane) direction, independent of the spin of the initial or final atom. By completing the sum in Eq.~(\ref{eq:hamtbs}) for the nearest neighbor terms in the $\hat{z}$-direction, or as used here, the interlayer direction, we find two contributions to the Hamiltonian. The first is
\begin{equation}
H_{inter} = \left[t_{z} \cos(k_{z})\cos(\frac{k_{x}}{2})\cos(\frac{k_{y}}{2})\right] \cdot (\Sigma_{R}^{0} + \Sigma_{L}^{0}),
\label{eq:hinter1_nn}
\end{equation}
with $t_{z}$ being the hopping in the interlayer direction. The second contribution to the interlayer nearest neighbor hopping Hamiltonian is
\begin{equation}
H_{intra} = i\left[t_{z} \sin(k_{z})\cos(\frac{k_{x}}{2})\cos(\frac{k_{y}}{2})\right] \cdot (\Sigma_{L}^{0} - \Sigma_{R}^{0}).
\label{eq:hinter2_nn}
\end{equation}

With a slight extension of this logic, we may include the next-nearest neighbor hopping term that  corresponds to non-spin dependent intercell hopping between Fe-Fe or Rh-Rh bonds. In order to define the array of matrix elements in the model in a general manner, we define the matrices
\begin{equation}
\Xi_{ij}^{k}=\left( \tau_{i}\otimes\tau_{j} \right)\otimes \sigma_{k},
\label{eq:gensig}
\end{equation}
where $\tau_{i}$ are the Pauli matrices used here to denote the occupied FeRh sites within the unit cell. Using this definition, we find the on-site next-nearest neighbor contribution as
\begin{equation}
H_{NNN} = [t_{xy}^{'}  (\cos(k_{x}) + \cos(k_{y})) + t_{xy}^{'} \cos(2k_{z})] \cdot \Xi_{00}^{0}.
\label{eq:honsite}
\end{equation}
In Eq.~(\ref{eq:honsite}), the additional $\cos(2k_{z})$ comes from the hopping from out of the top and bottom of the unit cell due to the separation between the length of the unit cell in the $\hat{z}$ or interlayer direction at it couples to other cells. Furthermore, there also needs to be a term included in the Hamiltonian that couples the $\langle{Fe_{1,\uparrow}}|$ and $\langle{Fe_{1,\downarrow}}|$ with the corresponding $|{Fe_{2,\uparrow}}\rangle$ and $|{Fe_{2,\downarrow}}\rangle$ and along with the Hermetian conjugate for this action. Additionally, the connection described for the Fe atoms must also be defined for the Rh atoms. We include the inter-cell coupling between next-nearest neighbor atoms in the interlayer direction using
\begin{equation}
H_{Cell} = t_{z}^{'} \cdot \Xi_{x0}^{0}.
\label{eq:hcellnnn}
\end{equation}
In addition to the hopping elements that we have outlined, the each of the non-magnetic hopping terms is altered by the presence of spin canting that results from the application of an in-plane magnetic field. In order to be able to account for this change, we utilize a model for double exchange\cite{anderson1955considerations} in which each of the non-magnetic hopping elements is multiplied by $\cos(\theta_{c}/2)$ where $\theta_{c}$ is the cant angle\cite{baublitz2014minimal}.

\subsubsection{Spin-Orbit Matrix Elements}

When FeRh is deposited or grown on a substrate, in this case MgO, the strain imparted into the FeRh layer is relaxed in the perpendicular direction and results in the Rh atoms no longer being constrained to sit in the center of the Fe atoms. The deviation of the Rh atomic positions within the  Fe lattice allows one to construct a spin-orbit interaction in an orthorhombic lattice structure using only the $s$-orbitals. In completing the sum shown in the second term of the real-space Hamiltonian, Eq.~($2$), the contributions to the spin-orbit coupling are parametrized by the coupling magnitude, $\lambda=\lambda_{xy}\times\left(1-0.2\cos(4\varphi)\right)$, where $\varphi$ is the angle of deviation between the crystal direction of the current flow direction and the $[100]$ crystal direction. The use of $\lambda(\varphi)$ that contains an explicit dependence on the crystal direction in which the current flows takes into account the assumed cubic magnetocrystaline anisotropy of FeRh and the details of our sample preparation. Furthermore, we use an approach in which the spin-orbit coupling is broken down into inter, $\lambda_{z}$, and intralayer, $\lambda_{xy}$, components so as to facilitate the ability to tune the interlayer coupling from strongly coupled to weakly coupled, in which the layers act more as 2D entities. Therefore, the interlayer spin-orbit coupling term may be written as $\lambda_{z}=\hat{\lambda_{z}}\lambda_{xy}$, where $\hat{\lambda_{z}}$ is a constant.

Using Eq.~(\ref{eq:gensig}), we may define the intralayer spin-orbit coupling term as,
\begin{equation}
H_{lxy} = \lambda_{xy} \cdot \left[ \Sigma_{0z}^{y} \sin(k_{x}) - \Xi_{0z}^{x} \sin(k_{y}) \right],
\label{eq:hso1}
\end{equation}
and the interlayer spin-orbit contribution as
\begin{equation}
H_{lz} = [\lambda_{z}\cos(k_{z})] \cdot \left[ \Xi_{0z}^{y} \sin(k_{x}) - \Xi_{0z}^{x} \sin(k_{y}) \right],
\label{eq:hso2}
\end{equation}

\subsubsection{Antiferromagnetic Exchange Interaction Matrix Element}

The interaction Hamiltonian is defined as
\begin{equation}
H_{int} = \sum_{i} J \mathbf{m_{i}} \cdot \mathbf{s_{i}} + \sum_{\langle ij \rangle} A_{ex} \mathbf{m_{i}} \cdot \mathbf{m_{j}},
\label{eq:Hint}
\end{equation}
where the spin-density operator is defined as $s_{i} = c_{i}^{\dagger} \mathbf{\sigma} c_{i}$. Under the mean-field approximation within a unit cell, a local magnetization $\mathbf{m_{i}}$ is assumed at the site $i$ with a uniform saturation magnetization, $\lvert \mathbf{m_{i}} \rvert = m_{s}$ for all lattice sites. In Eq.~(\ref{eq:Hint}), $J$ is the on-site exchange coupling constant between the itinerant electron spin $(s_{i})$ and the local magnetic moment $(m_{i})$, and $A_{ex}$ is the antiferromagnetic exchange constant between adjacent local magnetic moments.

We assume that the antiferromagnetic exchange strength is to satisfy the condition $m_{Fe}^{1} = -m_{Fe}^{2}$ and $m_{Rh}^{1} =  m_{Rh}^{2} = 0$. When an in-plane magnetic field is present, $m_{Rh}^{1} =  m_{Rh}^{2} \neq 0$. In this formulation, we define $J \mathbf{m}_{r}^{1} = \Delta^{r} \mathbf{\hat{n_{r}}}$, where the magnetization vector $\mathbf{\hat{n_{r}}}$ and the exchange magnitude $\Delta^{r}$ characterizes the orientation and the magnitude of the antiferromagnetic order parameter. As a result, the interacting Hamiltonian in Eq.~(\ref{eq:Hint}) becomes
\begin{equation}
H_{int} = \frac{1}{N} \sum_{r} \Delta^{r}\cdot \mathbf{\hat{n_{r}}} \cdot [\mathbf{s}_{r}^{1} - \mathbf{s}_{r}^{2}],
\label{eq:hint2}
\end{equation}
where $\mathbf{s}^{1/2}_{r}$ is a spin-density operator of the $1 / 2$ sublattice at the $r^{th}$ unit cell. The contribution to the local magnetic moments, or the second term in Eq.~(\ref{eq:Hint}) becomes constant for the assumed antiferromagnetic order by merely adding a constant shift in the total Hamiltonian, thus we may ignore its contribution to the interacting Hamiltonian in Eq.~(\ref{eq:hint2}).

In general, we assume within the mean-field approximation that local fluctuations in the magnetization are negligible. For this reason, we may approximate the local magnetic momentum by a global averaged value in Eq.~(\ref{eq:hint2}) and the magnetization vector, $\mathbf{\hat{n}}=[\cos(\theta)\cos(\phi),\sin(\phi)\sin(\theta),\sin(\phi)]$ where $\phi$ is the polar angle and $\theta$ is the azimuthal angle of the magnetization vector relative to the crystal direction of current flow. The antiferromagnetic exchange interaction within the FeRh tight-binding structure via \cite{wang2017antiferromagnetic,Kim2018} via
\begin{equation}
H_{afm} = H_{int} \cdot \Sigma_{\xi},
\label{eq:hafm}
\end{equation}
In Eq.~(\ref{eq:hafm}), we include the magnetization via 
\begin{equation}
\Sigma_{\xi} = \left[ (\tau_{z} + \tau_{0})/2 \right]\otimes \tau_0 \otimes \left[ \mathbf{\Vec{n}\cdot\mathbf{\Vec{\sigma}}} \right]\equiv \xi\otimes \left[ \mathbf{\Vec{n}\cdot\mathbf{\Vec{\sigma}}} \right],
\end{equation}
where $\xi$ is a diagonal $4\times4$ matrix that sets the weights of the magnetic order within the unit cell in order to keep the antiferromagnetic order between the Fe atoms and ferromagnetic order, as required when FeRh is immersed in an in-plane magnetic field, on the Rh atoms. 

The obvious conflict between the number of atoms contained within the unit cell of the tight-binding model and that of the DFT calculation requires additional consideration. In particular, we require that the additional symmetries that are present in the tight-binding model should not be reflected in the location of the resultant energy bands near the Fermi energy. Via the addition of symmetry breaking terms to the tight-binding model to break the in-plane and out-of-plane mirror, glide, and translational symmetries, we find that the calculated tight-binding bandstructure is altered at high energies, but remains relatively unchanged at energies near the Fermi level. In this manner, despite the presence of additional symmetries within the tight-binding model, the tight-binding and DFT representation of FeRh agree qualitatively at energies close to the Fermi surface, as seen in Supplemental Figure~\ref{fig:bandcomp}. 

\begin{figure*}
\begin{center}
\label{fig:bandcomp}
\includegraphics[width=12 cm]{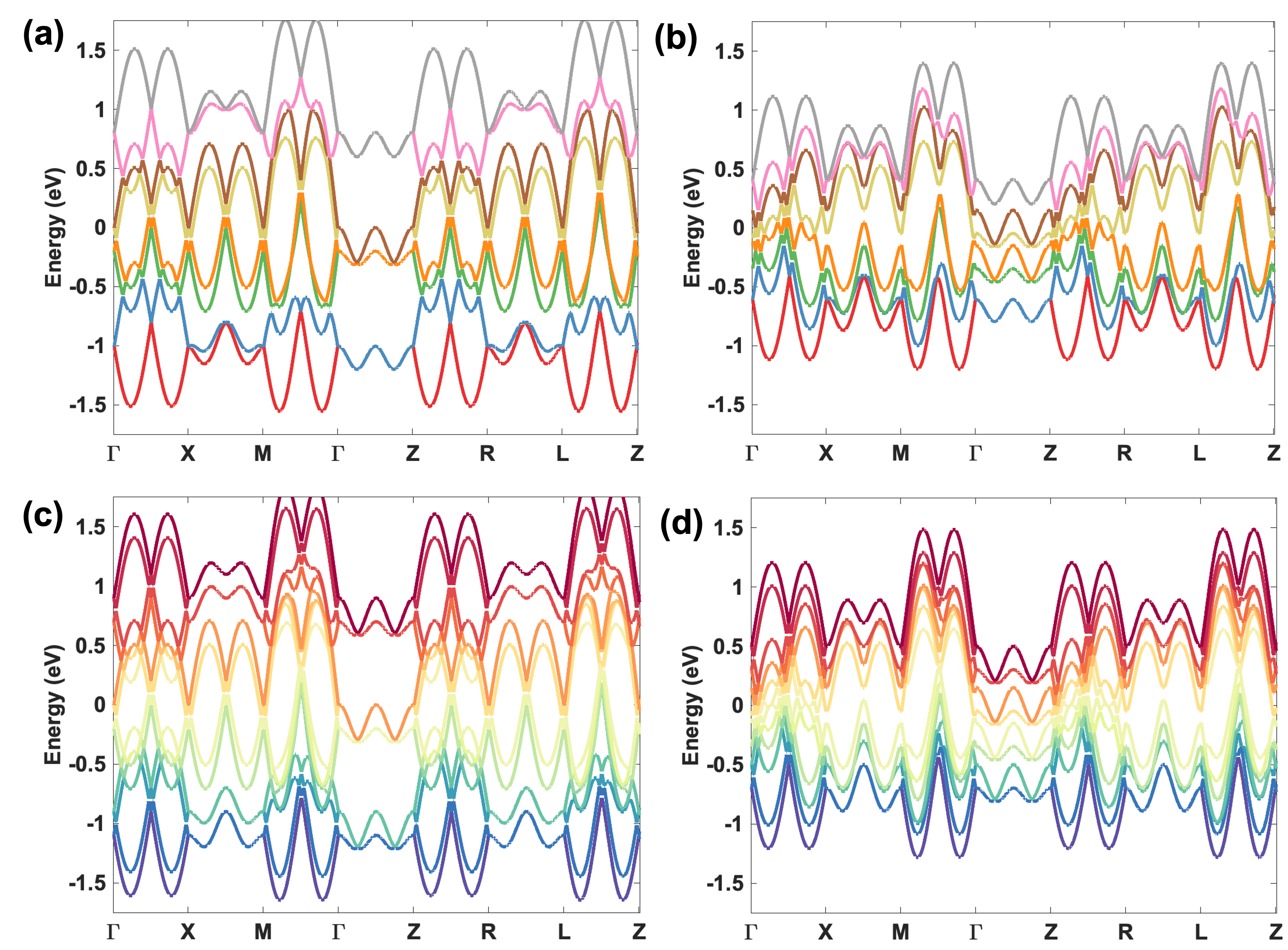}
\caption{\textbf{Comparison between $8$ and $16$ band tight-binding models of FeRh:} Bandstructure of $[100]$ oriented FeRh within the $8$ band representation calculated with $\mathbf{\vec{m}}$ pointing in the $[100]$ direction for a parameter choice of (a) $\Delta=0.9$, $M_{Rh}=0$, $\theta_{cant}=0^{\circ}$, $t_{xy}=0.3$, $t_{xy}^{'}=t_{z}^{'}=0.1$, $t_{z}=0.5$, $\lambda_{xy}=0.5$, and $\lambda_{z}=0.1$ showing the non-degenerate band touching points at and close to $E=0$ and (b) where the parameters $\Delta=0.5$, $M_{Rh}=0.15$, $\theta_{cant}=20^{\circ}$ have been modified to represent the bands past the Lifshitz transition with all other parameters unchanged. These are to be compared to the same parameter choices in (c) and (d) but represented in a $16$ band representation.}
\label{fig:tbbandcomp}
\end{center}
\end{figure*}

Under the application of an in-plane magnetic field, the initially inert Rh atoms will acquire a ferromagnetic moment that must be accounted for in each of the respective models. By accounting for the newly developed magnetic moment on the Rh, each of the models produces a reduced set of symmetry operations that are present. Within the framework of the DFT calculation, a magnetic moment is added to the Rh atom perpendicular to the canted Fe magnetic moments to approximate the physical situation under consideration in this work. The most obvious alteration in the models that results from the addition of the ferromagnetic moment on the Rh atoms is that the tight-binding model retains in-plane rotational and mirror symmetries that lead to the appearance of nodel lines within the Brillouin zone in the tight-binding model that are not present in the DFT. However, the presence of such nodal structure does not contribute to the quantum transport properties and, therefore, do not impact any of the results presented in the main text. 

\subsubsection{Quantum Transport in the Ferromagnetic Phase of FeRh and Ferromagnetic AMR in FeRh}
\label{sec:sup-qtfm}
\begin{figure*}
\begin{center}
\label{fig:dappl2}
\includegraphics[width=12 cm]{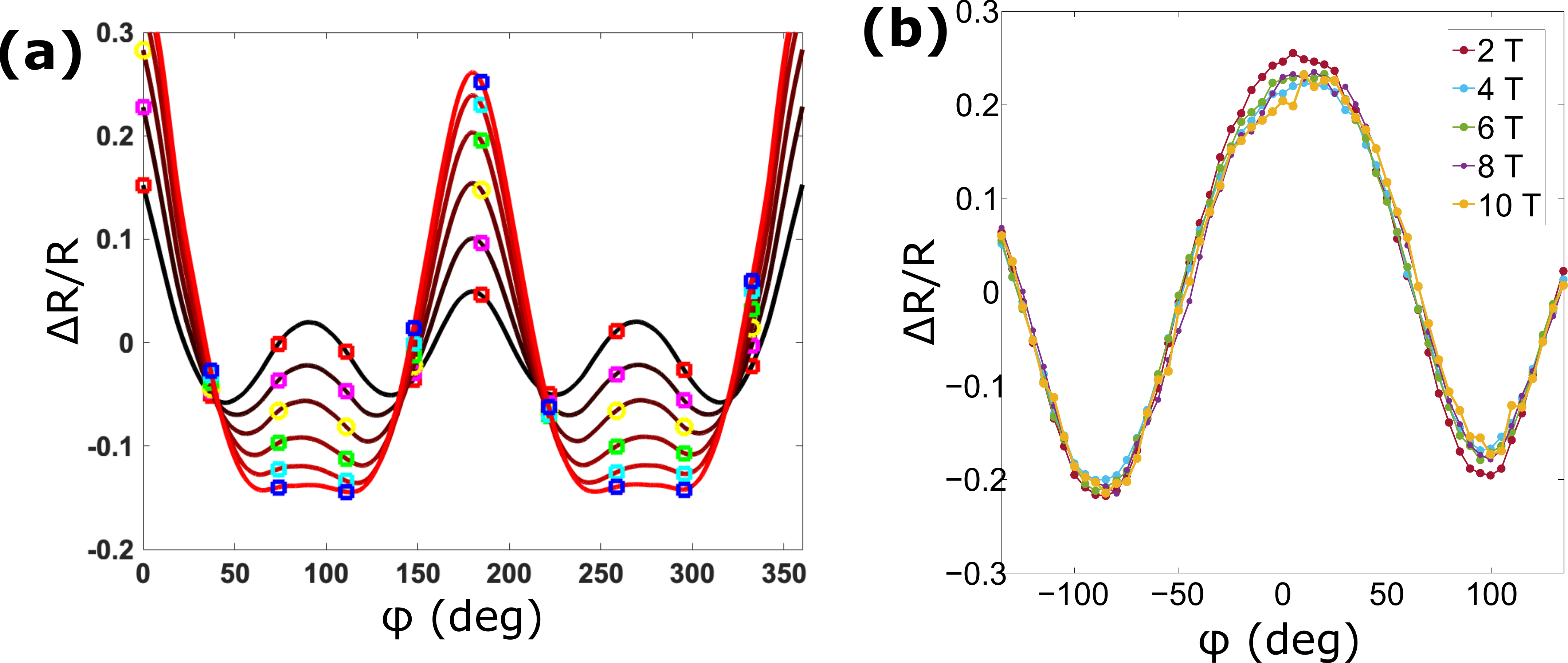}
\caption{\textbf{Anisotropic Magnetoresistance in Ferromagnetic $FeRh$:}(a) The change in resistance normalized by the average resistance along the $[100]$-direction as a function of the in-plane angle of the magnetic orientation with the current direction when $\Delta \geq \lambda$ within the ferromagnetic phase of FeRh. The magnitude of the critical parameters considered in this phase of the material are: $\Delta$ ranges from $0.9t$ (red curves) to $0.6t$ (black curves).  (b) The ferromagnetic AMR is measured at $310$~K for fields between $2$ - $10$~T. }
\label{fig:Tfmamr}
\end{center}
\end{figure*}
In order to further understand the validity of the theoretical model for thin-film FeRh, we examine the quantum transport properties of the ferromagnetic phase of FeRh. The transport properties of the ferromagnetic phase are fundamentally distinct from the antiferromagnetic phase, as examined extensively in the main text, in that there is no reduction in the magnetic order with the application of an in-plane magnetic field. Nonetheless, as the experiments show, we find that the magnetization orientation may be coherently manipulated by rotating the magnetic field. The resistance curves shown in Supplemental Figure~(\ref{fig:Tfmamr}) (a) show the same trends as in the AMR of the antiferromagnetic phase in that, as the magnitude of the ferromagnetic order decreases, the AMR does show signs of both $C_{2}$ and $C_{4}$-symmetric oscillations with the rotation of the field about the $[100]$, $\theta$, consistent with the $s-d$ scattering from the magnetic exchange field. We note that for the values of $\Delta$ and $\lambda$ that are selected here, we do not see the change in the sign of the AMR with increasing in-plane magnetic fields. However, using a lower magnitude of $\Delta$, thereby satisfying the conditions in the main text, will produce the anomalous behavior in the AMR.

The anisotropic magnetoresistance of the FeRh samples has  also been measured in the ferromagnetic phase.  The ferromagnetic AMR at $310$~K are shown in Supplemental Figure~(\ref{fig:Tfmamr}) (b).  Results are shown for a representative field range between $2$ - $10$~T.  Unlike the antiferromagnetic phase, the AMR exhibits no obvious dependence on the applied field magnitude.  This is expected because the coercive field of FeRh at this temperature is approximately $10$~mT.  Qualitatively, the AMR in the ferromagnetic phase shows a two-fold angular signal with the largest magnetoresistance occurring when the magnetization is oriented parallel to the external field.  

\subsection{Details of Quantum Transport Calculations}
\label{sec:sup-qttdets}

In this work, we make the connection between the transport experiments complete by calculating the dynamic response of the FeRh Hamiltonian within the linear response regime making extensive use of the non-equilibrium Green's function formalism\cite{anantram2008modeling,Datta1997}. The system temperature is set to be $T_{sys}=10$~K so as to include thermal broadening in accordance with the measurement base temperature. The transport calculation is designed to allow for current flow along the $[100]$ direction with the other two orthogonal directions kept in momentum space. Nonetheless, we must sum over the momentum contributions in the other two directions in momentum space to determine the total current flow, as
\begin{equation}
\begin{split}
J = \frac{e}{\hbar} \sum_{k_{\perp}} \int \frac{dE}{2\pi} ~ Re\left\{ Tr[t^{FeRh} G^{<}(\bf{k_{\perp}},E)] \right\}[f(\mu_{CL})-f(\mu_{CR})].
\end{split}
\label{eq:fcur}
\end{equation}
In Eq. (\ref{eq:fcur}), $\bf{k_{perp}} = (k_{y}-k_{z}) \in [-\pi \rightarrow \pi]$, $f(E)$ is the Fermi function, and $\mu_{CL}$ and $\mu_{CR}$ are the chemical potentials for the left and right contacts, respectively. Additionally, using the standard definitions of the Green's functions, $G^{<}$ is the lesser-than Green's function and $t^{FeRh}$ is the single particle next-nearest neighbor hopping element in the current flow direction of FeRh, which corresponds to transmission of charge between unit cells. The sum is taken over the range of energy that, naturally, depends on the applied bias between the ends of the system in question. As there are no dissipative mechanisms present in the system, then the relaxation will occur in the contact regions that inject and extract current from the system. We justify the lack of dissipative mechanisms within the simulations as an accurate representation of thin-film FeRh as, at $T=10~K$, (1) the disspative phonon contribution to the self-energy is expected to be small (2) disorder interactions renormalize the band energies but do not impart significant level broadening\cite{YKMJGPhysRevB98}. We note that were level broadening to be present, it would not impact the form of the AMR calculated theoretically, and (3) the inclusion of uncorrelated disorder, both concerning magnetic or non-magnetic impurities, within an appropriate self-energy term induces higher-order corrections to the Hamiltonian that do not alter transport dynamics\cite{PhysRevB.98.155430}.

Additionally, we may calculate the the spectral density of states (SDOS) that is defined as,
\begin{equation}
A_{spec} = SDOS = G^{r} \Gamma G^{a} = G^{a} \Gamma G^{r} = i[G^{r} - G^{a}].
\label{eq:dos}
\end{equation}
In Eq.~(\ref{eq:dos}), $G^{r}$ is the retarded Green's function, $G^{a}$ is the advanced Green's function, and $\Gamma$ represents the anti-Hermitian components of the corresponding self-energy terms as
\begin{equation}
\Gamma = i[\Sigma^{r} - \Sigma^{a}].
\end{equation}
In our case, we make a very simple approximation for the contact self-energies in which the real part of the self-energy term, associated with the energy shift of the levels, is neglected and only the level broadening effect is considered. The self-energy term also ignores the electronic structure of the metal leads and assumes a constant density of states and coupling constant from the metal to the system. The contact is an excellent approximation to the full surface Green's function when the density of states in the lead varies slowly as a function of energy. Under these conditions the self-energy term is deemed to be in the wide-band limit and the self-energy term is represented as,
\begin{equation}
\Sigma_{WBL} -i \frac{t_{c}^{2}\it{N(E_{F})}}{2}.
\label{eq:wbl}
\end{equation}
In Eq.~(\ref{eq:wbl}), $t_{c}$ is the hopping term that parametrizes hopping from the metal to the system and $\it{N(E_{F})}$ is the surface density of states at the Fermi energy, $E_{F}$. In the calculations, we assume that the the voltage across the central system is vanishingly small so that the transport at $E_{F}$ only needs to be considered. With the additional terms in the calculation defined, the number of real space points along the crystal direction where current flows and the number of $k$-points in $k_{y}$ and $k_{z}$ are sufficient to ensure that there are no quantization effects and that the results of the numerics do not change as a function of the number of points within the solution domain.

\subsection{Parameter Selection for FeRh Quantum Transport Model}
\label{sec:sup-paras}

\begin{table}[h!]
\begin{center}
 \begin{tabular}{||c | c | c | c ||}
 \hline
 \hline
 Parameter & Low $\theta_{cant}$ AMR & Intermediate $\theta_{cant}$ AMR & High $\theta_{cant}$ AMR \\ 
 \hline\hline
 \multirow{2}{6em}{$\Delta$} & \multirow{2}{6em}{$0.9$} & \multirow{2}{6em}{$0.62$} & \multirow{2}{6em}{$0.5$}\\
 & & & \\
 \hline
 \multirow{2}{6em}{$M_{Rh}$} & \multirow{2}{6em}{$0.0$} & \multirow{2}{6em}{$0.1$} & \multirow{2}{6em}{$0.15$}\\
 & & & \\
 \hline
 \multirow{2}{6em}{$\theta_{cant}$} & \multirow{2}{6em}{$0^{\circ}$} & \multirow{2}{6em}{$12.0^{\circ}$} & \multirow{2}{6em}{$14.5^{\circ}$}\\
 & & & \\
 \hline
 \multirow{2}{6em}{$t_{xy}$} & \multirow{2}{6em}{$0.3$} & \multirow{2}{6em}{$0.3$} & \multirow{2}{6em}{$0.3$}\\
 & & & \\
 \hline
 \multirow{2}{6em}{$t_{z}=t^{'}_{xy}=t^{'}_{z}$} & \multirow{2}{6em}{$0.1$} & \multirow{2}{6em}{$0.1$} & \multirow{2}{6em}{$0.1$}\\
 & & & \\
 \hline
 \multirow{2}{6em}{$\lambda^{[100]}_{xy}$} & 
 \multirow{2}{6em}{$0.5$} & \multirow{2}{6em}{$0.5$} & \multirow{2}{6em}{$0.5$}\\
 & & & \\
 \hline
 \multirow{2}{6em}{$\hat{\lambda_{z}}$} & 
 \multirow{2}{6em}{$1$} & \multirow{2}{6em}{$1$} & \multirow{2}{6em}{$1$}\\
 & & & \\
 \hline\hline
\end{tabular}
\caption{Complete list of the values utilized in the parameters of the tight-binding model for FeRh. Unless otherwise noted, all of the units are in $eV$. The value of $\lambda_{xy}$ is given along the $[100]$ crystal direction in FeRh and changes when the current flow is directed along another crystal axis. $\hat{\lambda_{z}}$ is chosen to indicate that there are no modifications to the spin-orbit coupling in the interlayer, $\hat{z}$, direction. }\label{tab:Sparms}
\end{center}
\end{table}
In Supplementary Table~\ref{tab:Sparms}, we provide all of the values corresponding to the parameters we use in the tight-binding model of FeRh. The values that are given for the points where we observe the transitions from one phase to the next. The goal of the tight-binding model is to qualitatively demonstrate the physical phenomena present in FeRh and, as such, does not include many other factors that are likely to cause the qualitative discrepancies such as: scattering, disorder, contact effects, strain, and the ferromagnetic substrate, to name several obvious omissions. Within the NEGF simulations, there is no explicit inclusion of a magnetic field as the experimental measurements see no orbital effects leading to the use of a $1$D transport model. Therefore, the key magnetic parameters ($\Delta$, $M_{Rh}$, and $\theta_{cant}$) are varied linearly in the model predicated on density function theory and other tight-binding calculations\cite{zarkevich2018ferh,khan1981origin,Hasegawa1987,Koenig1982} on FeRh.

\subsection{Fermi Surface Morphology Evolution}
\label{sec:sup-dosfs}
\begin{figure}
\begin{center}
\includegraphics[width=14cm, height=8cm]{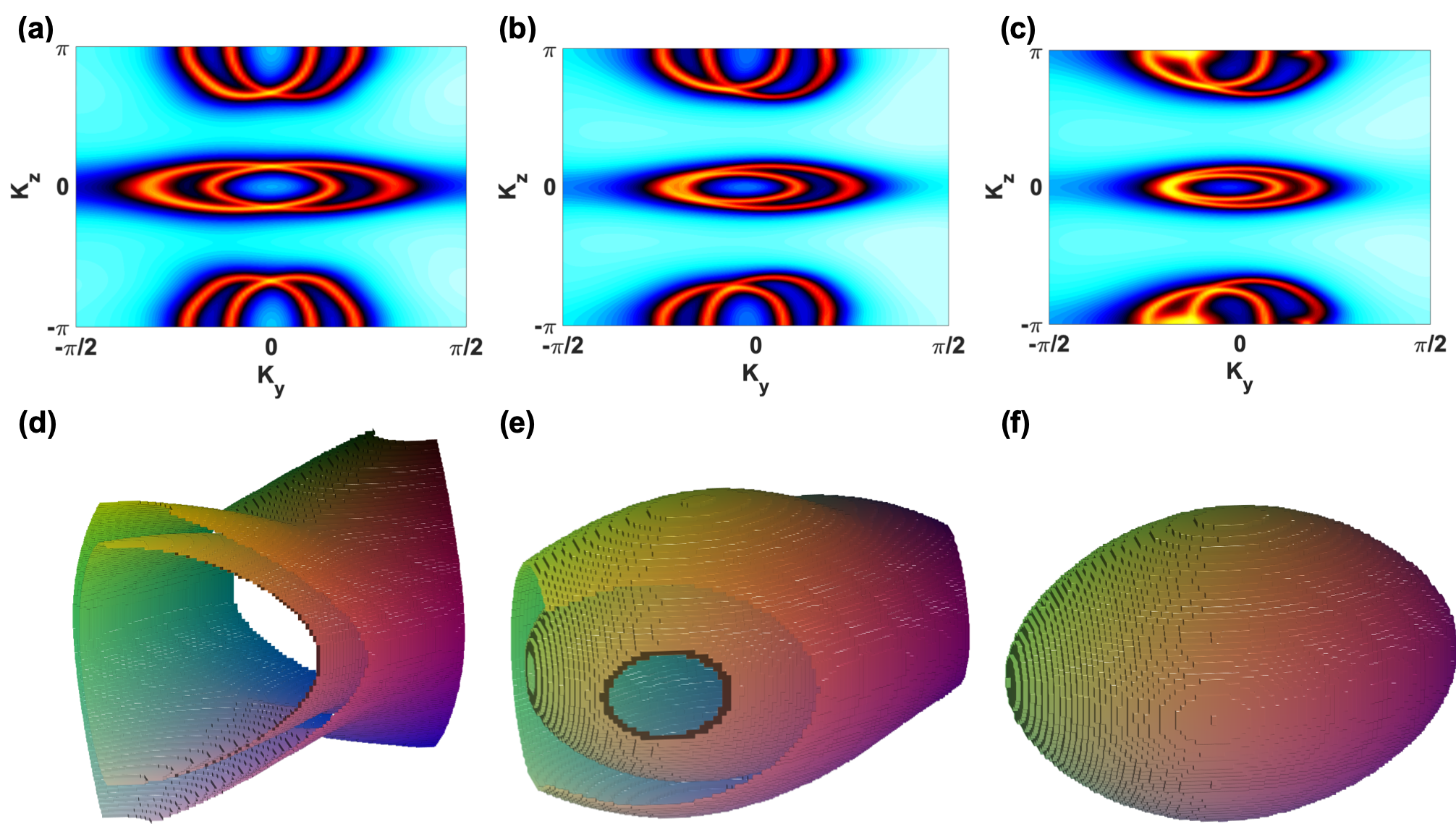}
\caption{Fermi Surface of FeRh: Local density of states calculated at $E_{F}=0$ when (a) $\Delta \gg \lambda$ (b) $\Delta \approx \lambda$ and (c) $\Delta \leq \lambda$ focused on the area of $k_{y}\in\{-\frac{\pi}{2},\frac{\pi}{2}\}$ and $k_{z}\in\{-\pi,\pi\}$ at $k_{x}=0$ corresponding to net magnetization direction $\phi=45^{\circ}$. We note that as the AFM exchange term is reduced in value compared to the spin-orbit coupling term, the central nodal structure undergoes a Lifshitz transition from and surface becomes more elongated in momentum space as a result of the coupling to the pseudogravitational fields. Fermi surface plots of the central nodal structure considered in (a)-(c) for (d)  $\Delta \gg \lambda$ (e) $\Delta \approx \lambda$ and (f) $\Delta \leq \lambda$ corresponding to net magnetization direction $\phi=45^{\circ}$. The evolution of the Fermi surface shows the transition and the change in structure as the parameters of the model are changed. }
\label{fig:DOSFS}
\end{center}
\end{figure}
In the main text, we argue that the changes in the Fermi surface are correlated with the coupling to the background pseudogravitational fields. In Supplemental Figure~\ref{fig:DOSFS}, we show the evolution to the Fermi surface when the corresponding net magnetization direction is $\phi=45^{\circ}$ as a function of the antiferromagnetic exchange in the $k_{y}-k_{z}$ plane at $k_{x}=0$. In Supplemental Figure~\ref{fig:DOSFS}(a) we show the SDOS corresponding to the case when $\Delta \gg \lambda$ at $E_{F}=0$. In Supplemental Figure~\ref{fig:DOSFS}(d), we plot the corresponding $3$D surface about the Weyl nodes near, but not located exactly upon, the $\Gamma$ point. We notice that the shape of the surface extends in response to the presence of remaining nodal structure in the $k_{x}-k_{y}$ plane. As $\Delta$ is reduced in to $\Delta \approx \lambda$ in Supplemental Figure~\ref{fig:DOSFS}(b), the arcs on the $k_{z}=\pm \pi$ surface have shifted and become more elongated in momentum space in a similar fashion to the nodes located at $k_{z}=\pm \pi$. We attribute the elongation to the coupling to the background pseudogravitational fields resulting in the observed distortion. In Supplemental Figure~\ref{fig:DOSFS}(e), we observe that the central nodes have also shifted and elongated in momentum space extending farther in $k_{y}$ with reduced presence in $k_{z}$. As parameter values are further decreased to satisfy $\Delta \leq \lambda$ in Supplemental Figure~\ref{fig:DOSFS}(c), the region in which the pseudogravitational fields dominate the charge transport, we observe further distortions to the Fermi surface evidenced by the significantly warped SDOS connecting the Weyl points within the momentum cut. It is notable that given the off-axis orientation of $\mathbf{\vec{m}}$ in Supplemental Figure~\ref{fig:DOSFS}(c) indicates that there is an overall additional presence of states on $-k_{y}$ that increases with decreasing $\Delta$ corresponding to the increased importance of the pseudogravitational coupling in the morphology of the constituent Fermi surfaces. The distorted individual Fermi surfaces near the center of the momentum space and the Lifshitz transition has completed with one Fermi surface nested inside of the other while both have been reduced in size. In Supplemental Figure~\ref{fig:DOSFS}(f), the complete nesting of the Fermi surfaces are observed in the $3$D rendering as we are only able to observe the a single distorted Fermi surface as compared to the renderings in Supplemental Figures~\ref{fig:DOSFS}(c) and ~\ref{fig:DOSFS}(d).

\bibliographystyle{unsrt}
\section*{Supplementary References}
\bibliography{Weyl}